\documentclass[lettersize,journal]{IEEEtran}
\usepackage{amsmath,amsfonts}
\usepackage{xcolor}
\usepackage{algorithmic}
\usepackage{algorithm}
\usepackage{array}
\usepackage[caption=false,font=normalsize,labelfont=sf,textfont=sf]{subfig}
\usepackage{textcomp}
\usepackage{stfloats}
\usepackage{url}
\usepackage{verbatim}
\usepackage{graphicx}
\usepackage{cite}

\newcommand {\Define} {\stackrel {\Delta} {=}  }

\newtheorem{theorem}{Theorem}

\begin{document}

\title{Zak-OTFS with Interleaved Pilots to Extend the Region of Predictable Operation}

\author{Jinu Jayachandran, Imran Ali Khan, Saif Khan Mohammed, Ronny Hadani,
Ananthanarayanan Chockalingam and Robert Calderbank, ~\IEEEmembership{Fellow,~IEEE}
\thanks{J. Jayachandran, I. A. Khan  and S. K. Mohammed are with Department of Electrical Engineering, Indian Institute of Technology Delhi, India (E-mail: Jinu.Jayachandran@ee.iitd.ac.in, Imran.Ali.Khan@ee.iitd.ac.in, saifkmohammed@gmail.com). S. K. Mohammed is also associated with Bharti School of Telecom. Technology and Management (BSTTM), IIT Delhi. R. Hadani is with Department of Mathematics, University of Texas at Austin, TX, USA (E-mail: hadani@math.utexas.edu). A. Chockalingam is with Department of Electrical Communication Engineering, Indian Institute of Science Bangalore, India (E-mail: achockal@iisc.ac.in). R. Calderbank is with Department of Electrical and Computer Engineering, Duke University, USA (E-mail: robert.calderbank@duke.edu).}

\thanks{The work of Saif Khan Mohammed was
supported in part by a project (at BSTTM) sponsored by Bharti Airtel Limited, and in part by the Jai Gupta Chair at IIT Delhi.}

\thanks{\textbf{This work has been submitted to the IEEE for possible publication.
Copyright may be transferred without notice, after which this version may
no longer be accessible.}}
}



\maketitle

\begin{abstract}
When the delay period of the Zak-OTFS carrier is greater than the delay spread of the channel, and the Doppler period of the carrier is greater than the Doppler spread of the channel, the effective channel filter taps can simply be read off from the response to a single pilot carrier waveform. The input-output (I/O) relation can then be reconstructed for a sampled system that operates under finite duration and bandwidth constraints. We introduce a framework for pilot design in the delay-Doppler (DD) domain which makes it possible to support users with very different delay-Doppler characteristics when it is not possible to choose a single delay and Doppler period to support all users. {The method is to
interleave single pilots in the DD domain, and to choose the
pilot spacing so that the I/O relation can be reconstructed
by solving a small linear system of equations.}
\end{abstract}

\begin{IEEEkeywords}
Zak-OTFS, predictable, pilot, interleaved, DD domain.
\end{IEEEkeywords}

\section{Introduction}
The Zak-OTFS carrier waveform is a pulse in the delay-Doppler(DD) domain, that is a quasi-periodic localized function defined by a delay period $\tau_p$ and a Doppler period $\nu_p$ where $\tau_p \, \nu_p = 1$. The time-domain (TD) realization of the carrier is a pulsone, that is a train of pulses modulated by a tone where adjacent pulses are spaced $\tau_p$ seconds apart. The frequency-domain (FD) realization of the carrier is a train of pulses in the frequency domain (FD) modulated by a FD sinusoid where adjacent pulses are spaced $\nu_p$ Hz apart. We have shown \cite{zakotfs2} that the Zak-OTFS input-output (I/O) relation is predictable\footnote{\footnotesize{Predictability implies that the channel response to an input DD pulse at any arbitrary discrete DD location $(k_p, l_p)$ can be predicted from the knowledge of the channel response to a DD pulse at some other location.}} and non-fading\footnote{\footnotesize{Consider the DD domain energy distribution of the channel response to an input DD pulse. The I/O relation is said to be non-fading if the energy distribution around the pulse is invariant of its location.}} when the delay period of the pulsone is greater than the delay spread of the channel and the Doppler period of the pulsone is greater than the Doppler spread of the channel \cite{otfsbook}. We refer to this condition as the crystallization condition. When the crystallization condition holds, the taps of the effective DD domain channel filter can simply be read off from the DD domain response to a single pilot carrier waveform, and the I/O relation can be reconstructed for a sampled system that operates under finite duration and bandwidth constraints. Section \ref{secsysmodel} describes the Zak-OTFS system model.

4G and 5G wireless communication networks use OFDM rather than Zak-OTFS. However OFDM exhibits poor reliability for high delay and Doppler spreads characteristic of next generation communication scenarios \cite{Wang2006}, \cite{IMT2030, intro1, intro2}. The first instantiation of OTFS  was designed to be compatible
with 4G/5G modems and is called MC-OTFS (Multicarrier OTFS).
MC-OTFS is superior to OFDM for high delay and Doppler spreads
\cite{mcotfs1, embed1}, but is inferior to Zak-OTFS \cite{zakotfs2}. Note that in MC-OTFS, modulation, detection and estimation are all performed in the DD domain \cite{mcotfs2, channel, OTFSMP,embed2}. The OTFS Special Interest Group (SIG) website \cite{mcotfs3} is a rich source of information about MC-OTFS.


In MC-OTFS, DD domain information symbols are transformed
to time-frequency (TF) symbols which are then used to generate
the transmitted TD signal. This two-step modulation can be avoided by using the Zak-transform \cite{Zak67, Janssen88}
to obtain the transmitted TD signal directly from the DD domain
information symbols (see \cite{derotfs, zak3} for details). This method of modulation is called Zak-OTFS (see \cite{zakotfs1, zakotfs2} for details) and it achieves better throughput/reliability than MC-OTFS, particularly in high delay/Doppler spread scenarios. There are also implementations based on the discrete Zak transform \cite{Lampel22} and on TF windowing \cite{Hanly23}. Filtering in the discrete DD domain
generate noise-like pilot waveforms (spread pilots) that enable integrated sensing and communication (ISAC) in the same Zak-OTFS subframe \cite{zakotfs3}.

We emphasize that we are learning the I/O relation without estimating the physical channel parameters (gain, delay and Doppler shift of each physical path). By focusing not on acquiring the channel, but on acquiring the interaction of channel and modulation, Zak-OTFS circumvents the legacy channel model dependent approach to wireless communications and operates model-free. We present numerical simulations in Section \ref{simsec} for the Veh-A channel model \cite{EVAITU} which consists of six channel paths and is representative of real propagation environments. In these simulations, we deliberately choose the channel bandwidth so that not all paths are separable/resolvable, and it is not possible to estimate the physical channel.

Why Zak-OTFS rather than OFDM? Perhaps the most important reason is that 6G propagation environments are changing the balance between time-frequency methods focused on OFDM signal processing and delay-Doppler (DD) methods focused on Zak-OTFS signal processing. OFDM signals live on a coarse information grid (i.e., integer multiples of the sub-carrier spacing), and cyclic prefix/carrier spacing are designed to prevent inter carrier interference (ICI). When there is no ICI, equalization in OFDM is relatively simple once the I/O relation is acquired. However, acquisition of the Input/Output (I/O) relation in OFDM is non-trivial and model-dependent, and the interaction of the OFDM carrier with the channel varies in both TD and FD. By contrast Zak-OTFS signals live on a fine information grid.\footnote{\footnotesize{In Zak-OTFS, information symbols are carried by DD pulses having locations separated by integer multiples of the inverse bandwidth along the delay domain and separated by integer multiples of the inverse subframe duration along the Doppler domain.}} Since information carrying DD pulses are located on a finer grid, they interfere with each other resulting in inter-carrier interference due to which equalization is more involved. However, when the crystallization condition holds the I/O relation can be read off from the response to a single pilot signal. In this paper we consider acquiring the I/O relation.
\begin{figure}
\vspace{-7mm}
\centering
\includegraphics[width=8.5cm, height=6.5cm]{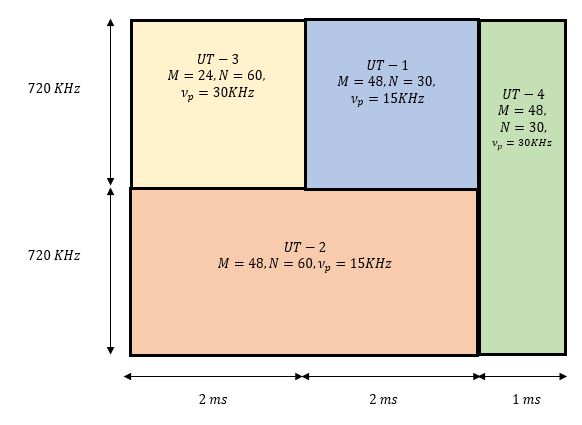}
\caption{{Orthogonal/non-overlapping allocation of time-frequency (TF) resource to different users having distinct delay-Doppler profile.}} 
\label{figutalloc}
\end{figure}

{In a time and bandwidth limited communication system, users can be allocated different non-overlapping time-frequency (TF) resources by simply shifting each user's signal in time and frequency (orthogonal multiple-access).
In this paper, we consider Zak-OTFS based orthogonal TF access (there is no multi-user interference). A typical orthogonal allocation for four user terminals (UTs) is illustrated in Fig.~\ref{figutalloc}.
One important issue with OFDM, is that its numerology (sub-carrier spacing) is not flexible. In OFDM, a single user subject to inter-carrier interference causes carrier spacing to increase for all users, since OFDM does not allow for users to have different sub-carrier spacing. This increases the overall cyclic-prefix (CP) overhead substantially since a large sub-carrier spacing
implies a small OFDM symbol duration whereas the CP duration is determined by the delay-spread of the channel.}

{On the other hand, Zak-OTFS numerology is more flexible. Each user can choose a different delay/Doppler period parameter $(\tau_p, \nu_p)$ depending on its own channel, i.e., a user which experiences a higher Doppler spread would choose a higher $\nu_p$ so that the crystallization condition holds (in Fig.~\ref{figutalloc}, UT3 and UT4 experience higher Doppler spreads than UT1 and UT2). However, the implementation of such a system with different delay/Doppler period parameters for different users is challenging.
Therefore, in this paper we consider supporting users with different delay-Doppler characteristics without changing the delay and Doppler periods of their Zak-OTFS modulation. We propose interleaved DD domain pilots, so that with $Q$ interleaved pilots it is possible to accurately acquire the I/O relation as long as the Doppler spread $2 \nu_{max}$ is less than $Q \nu_p$. Although the associated pilot/guard overhead is higher for larger number of interleaved pilots, each user separately configures its own number of interleaved pilots depending on only its own channel Doppler spread.}

Section \ref{sectwopilot} describes how to place two \emph{interleaved} pilots on the original Zak-OTFS grid so that the I/O relation for the second user can be obtained by solving a $2 \times 2$ linear system. The method is simple and effective, but it is not the maximum likelihood (ML) estimate. In Section \ref{autoambig}, we analyze the ML estimator which is given by the samples of the cross-ambiguity between the received DD domain interleaved pilot and the transmitted DD domain interleaved pilot. The cross-ambiguity function is supported on a rectangular lattice in the DD domain, and the effective channel taps can be read off by restricting to any fundamental domain of this lattice. The delay and Doppler spacing of this lattice determine a second effective crystallization condition and interleaved pilots make it possible to support users that satisfy either of the two crystallization conditions on a system with a single delay period and single Doppler period.

{In Section \ref{simsec} we simulate the bit error rate (BER) performance of a Zak-OTFS subframe with interleaved pilots. For a fixed data signal power to noise power ratio (SNR) and fixed pilot power to data power ratio (PDR), it is observed that with every doubling in the number of interleaved pilots the maximum Doppler spread for which reliable/predictable operation is achieved, is also roughly doubled, i.e., \emph{extension} in the region of predictable operation. Also, the peak to average power ratio (PAPR) of the transmit TD signal reduces by $3$ dB for every doubling in the number of interleaved pilots.}

\section{System model}
\label{secsysmodel}
\begin{figure*}
\vspace{-7mm}
\centering
\includegraphics[width=16.5cm, height=4.5cm]{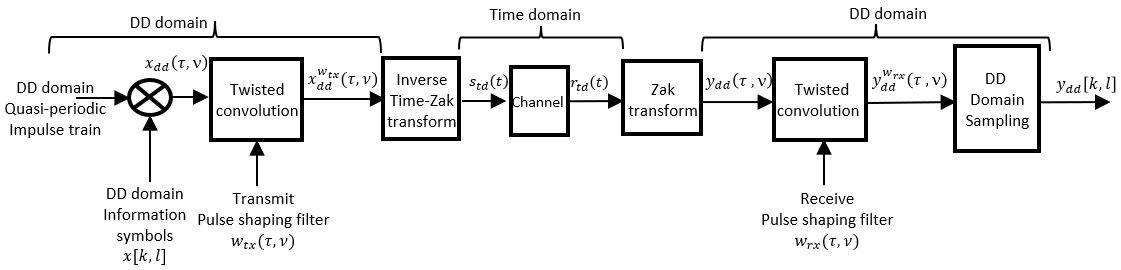}
\caption{Zak-OTFS transceiver processing.} 
\label{figzakotfspaper2}
\end{figure*}

Zak-OTFS transceiver processing is illustrated in Fig.~\ref{figzakotfspaper2} (see Section II of \cite{zakotfs2} also).
We transmit $MN$ symbols $x[k_0,l_0]$ in each subframe, $k_0=0,1,\cdots, M-1$, $l_0=0,1,\cdots, N-1$. The discrete DD domain pulse $x_{\mbox{\scriptsize{dd}}}^{(k_0, l_0)}[k,l]$ carries the information symbol $x[k_0, l_0]$, i.e.
 \begin{eqnarray}
 \label{eqn3p4}
     x_{\mbox{\scriptsize{dd}}}^{(k_0, l_0)}[k,l] & \hspace{-3mm} = &  \hspace{-3mm} \sum\limits_{n,m \in {\mathbb Z}} \hspace{-2mm} {\Big (} e^{j 2 \pi \frac{ n  l_0}{N}} \, x[k_0  , l_0] \, \delta[k - k_0 - nM] \nonumber \\
     & & \hspace{10mm} \delta[l - l_0 - mN]{\Big )},
 \end{eqnarray}for all $k,l \in {\mathbb Z}$. From (\ref{eqn3p4}) it is clear that this pulse carrying the $(k_0, l_0)$-th information symbol consists of infinitely many Dirac-delta impulses at discrete DD locations $(k_0 + nM, l_0 + mN)$, $n,m \in {\mathbb Z}$. Also, irrespective of $(k_0, l_0)$, $x_{\mbox{\scriptsize{dd}}}^{(k_0, l_0)}[k,l]$ is a \emph{quasi-periodic} function with period $M$ along the delay axis and period $N$ along the Doppler axis, i.e., for any $n,m \in {\mathbb Z}$, $x_{\mbox{\scriptsize{dd}}}^{(k_0, l_0)}[k,l]$ satisfies  
 \begin{eqnarray}
     \label{qpeqn}
     x_{\mbox{\scriptsize{dd}}}^{(k_0, l_0)}[k + nM,l + mN] & = & e^{j 2 \pi \frac{n l_0}{N}} \, x_{\mbox{\scriptsize{dd}}}^{(k_0, l_0)}[k,l].
 \end{eqnarray}The discrete DD domain pulses corresponding to all $MN$ information symbols are superimposed resulting in the discrete quasi-periodic DD domain signal
 \begin{eqnarray}
     x_{\mbox{\scriptsize{dd}}}[k,l] & = & \sum\limits_{k_0=0}^{M-1} \sum\limits_{l_0 = 0}^{N-1} x_{\mbox{\scriptsize{dd}}}^{(k_0, l_0)}[k,l].
 \end{eqnarray}
 $x_{\mbox{\scriptsize{dd}}}[k,l]$ is supported on the \emph{information lattice} $\Lambda_{dd} = \{\left( \frac{k \tau_p}{M} , \frac{l \nu_p}{N} \right) \, | \, k, l \in {\mathbb Z} \}$, i.e., we lift the discrete signal $x_{\mbox{\scriptsize{dd}}}[k,l]$ to the continuous DD domain signal
 \begin{eqnarray}
 \label{eqn2}
     x_{\mbox{\scriptsize{dd}}}(\tau, \nu) & = & \sum\limits_{k,l \in {\mathbb Z}} x_{\mbox{\scriptsize{dd}}}[k,l] \, \delta\left(\tau -  \frac{k \tau_p}{M} \right) \, \delta\left(\nu -  \frac{l \nu_p}{N} \right).
 \end{eqnarray}Note that, for any $n,m \in {\mathbb Z}$
 \begin{eqnarray}
     x_{\mbox{\scriptsize{dd}}}(\tau + n \tau_p, \nu + m \nu_p) & = & e^{j 2 \pi n \nu \tau_p} \, x_{\mbox{\scriptsize{dd}}}(\tau, \nu),
 \end{eqnarray}so that $x_{\mbox{\scriptsize{dd}}}(\tau, \nu)$ is periodic with period $\nu_p$ along the Doppler axis and quasi-periodic with period $\tau_p$ along the delay axis. 

 We use a pulse shaping filter $w_{tx}(\tau, \nu)$
 to limit the TD Zak-OTFS subframe to time duration $T = N \tau_p$ and bandwidth $B = M \nu_p$.
  The DD domain transmit signal $x_{\mbox{\scriptsize{dd}}}^{w_{tx}}(\tau, \nu)$ is given by the twisted convolution\footnote{\footnotesize{For any two DD functions, $a(\tau, \nu), b(\tau, \nu)$,  $a(\tau, \nu) *_{\sigma} b(\tau, \nu) = \iint a(\tau', \nu') \, b(\tau - \tau', \nu - \nu') \, e^{j 2 \pi \nu'(\tau - \tau')} \, d\tau' \, d\nu' $.}} of the transmit pulse shaping filter $w_{tx}(\tau, \nu)$
  with $x_{\mbox{\scriptsize{dd}}}(\tau, \nu)$. 
 \begin{eqnarray}
 \label{eqn61094}
     x_{\mbox{\scriptsize{dd}}}^{w_{tx}}(\tau, \nu) & = & w_{tx}(\tau, \nu) \, *_{\sigma} \, x_{\mbox{\scriptsize{dd}}}(\tau, \nu),
 \end{eqnarray}where $*_{\sigma}$ denotes the twisted convolution operator \cite{zakotfs1, zakotfs2}.
 The TD realization of $x_{\mbox{\scriptsize{dd}}}^{w_{tx}}(\tau, \nu)$
 gives the transmitted TD signal which is given by
 \begin{eqnarray}
 \label{eqn8p4}
 s_{\mbox{\scriptsize{td}}}(t) & = & {\mathcal Z}_t^{-1}\left( x_{\mbox{\scriptsize{dd}}}^{w_{tx}}(\tau, \nu) \right),
 \end{eqnarray}where ${\mathcal Z}_t^{-1}$ denotes the inverse Zak transform (see Eqn.~$(7)$ in \cite{zakotfs1} for more details).\footnote{\footnotesize{Just as the Fourier transform relates the TD and FD realizations of a signal, the Zak transform relates the TD and DD realizations of a signal. The inverse Zak-transform of a quasi-periodic continuous DD domain function/signal gives its TD realization and the Zak-transform of a TD signal gives its DD realization. Note that TD realization only exists for \emph{quasi-periodic} DD domain functions. Since twisted convolution of a quasi-periodic DD function with any arbitrary DD function is quasi-periodic, it follows that $x_{\mbox{\scriptsize{dd}}}^{w_{tx}}(\tau, \nu)$ in (\ref{eqn61094}) is also quasi-periodic.}}

The received TD signal is given by
\begin{eqnarray}
\label{eqn712446}
    r_{\mbox{\scriptsize{td}}}(t) & \hspace{-3mm} = & \hspace{-3mm} \iint h_{\mbox{\scriptsize{phy}}}(\tau, \nu) \, s_{\mbox{\scriptsize{td}}}(t - \tau) \, e^{j 2 \pi \nu (t - \tau)} \, d\tau \, d\nu \, + \, n_{\mbox{\scriptsize{td}}}(t),\nonumber \\
\end{eqnarray}where $h_{\mbox{\scriptsize{phy}}}(\tau, \nu)$ is the delay-Doppler spreading function of the physical channel and $n_{\mbox{\scriptsize{td}}}(t)$ is AWGN.
At the receiver, we pass from the TD to the DD domain by applying the Zak transform ${\mathcal Z}_t$ to the received TD signal $r_{\mbox{\scriptsize{td}}}(t)$, and we obtain
\begin{eqnarray}
\label{eqn712448}
    y_{\mbox{\scriptsize{dd}}}(\tau, \nu) & = & {\mathcal Z}_t\left( r_{\mbox{\scriptsize{td}}}(t) \right).
\end{eqnarray}Substituting (\ref{eqn712446}) into (\ref{eqn712448}) it follows that \cite{zakotfs1, zakotfs2}
\begin{eqnarray}
    y_{\mbox{\scriptsize{dd}}}(\tau, \nu) & = &  h_{\mbox{\scriptsize{phy}}}(\tau, \nu) \, *_{\sigma} \, x_{\mbox{\scriptsize{dd}}}^{w_{tx}}(\tau, \nu) \, + \, n_{\mbox{\scriptsize{dd}}}(\tau, \nu),
\end{eqnarray}where $n_{\mbox{\scriptsize{dd}}}(\tau, \nu)$ is the DD representation of the AWGN. Note that, in the DD domain the channel acts on the input $x_{\mbox{\scriptsize{dd}}}^{w_{tx}}(\tau, \nu)$ through twisted convolution with $h_{\mbox{\scriptsize{phy}}}(\tau, \nu)$. This is similar to how in linear time invariant (LTI) channels (i.e., delay-only channels), the channel acts on a TD input through linear convolution with the TD channel impulse response.
Twisted convolution is the generalization of linear convolution for doubly-spread channels. 

Next, we apply a matched filter $w_{rx}(\tau, \nu)$ which acts by twisted convolution on $y_{\mbox{\scriptsize{dd}}}(\tau, \nu)$ to give
\begin{eqnarray}
    y_{\mbox{\scriptsize{dd}}}^{w_{rx}}(\tau, \nu) & = & w_{rx}(\tau, \nu) *_{\sigma} y_{\mbox{\scriptsize{dd}}}(\tau, \nu)
\end{eqnarray}This filtered signal is then sampled on the information lattice $\Lambda_{\mbox{\scriptsize{dd}}}$ resulting in the quasi-periodic discrete DD domain signal $y_{\mbox{\scriptsize{dd}}}[k,l]$.
This discrete DD output signal is related to the input discrete DD signal
$x_{\mbox{\scriptsize{dd}}}[k,l]$ through the input-output (I/O) relation\footnote{\footnotesize{{The relation between discrete and continuous I/O relation for delay-only channels and doubly-spread channels is similar and has been described in detail in \cite{zakotfs1}
  (see Tables-I, II and III and the related discussion in \cite{zakotfs1}). In delay-only channels, the discrete-time I/O relation forms the basis for practical implementation and this is also the case for doubly-spread channels where the discrete-DD domain I/O relation (see (\ref{eqnio1})) forms the basis for practical implementation.}}} \cite{zakotfs1, zakotfs2}
\begin{eqnarray}
\label{eqnio1}
    y_{\mbox{\scriptsize{dd}}}[k,l] & = & h_{\mbox{\scriptsize{eff}}}[k,l] \, *_{\sigma} \, x_{\mbox{\scriptsize{dd}}}[k,l] \, + \, n_{\mbox{\scriptsize{dd}}}[k,l],
\end{eqnarray}where\footnote{\footnotesize{For any two discrete DD functions $a[k,l]$ and $b[k,l]$, the discrete twisted convolution between them i.e., $a[k,l] *_{\sigma} b[k,l] = \sum\limits_{k',l' \in {\mathbb Z}} a[k', l'] \, b[k - k', l- l'] \, e^{j 2 \pi l' \frac{(k - k')}{MN}}$.}} $h_{\mbox{\scriptsize{eff}}}[k,l]$ is the effective DD domain channel filter and $n_{\mbox{\scriptsize{dd}}}[k,l]$ are the DD domain noise samples. Note that $h_{\mbox{\scriptsize{eff}}}k,l]$ is simply   
\begin{eqnarray}
    h_{\mbox{\scriptsize{eff}}}(\tau,\nu) = w_{rx}(\tau, \nu) *_{\sigma} h_{\mbox{\scriptsize{phy}}}(\tau, \nu) *_{\sigma} w_{tx}(\tau, \nu) 
\end{eqnarray}sampled on the information lattice $\Lambda_{\mbox{\scriptsize{dd}}}$, i.e.
\begin{eqnarray}
    h_{\mbox{\scriptsize{eff}}}[k,l] & \Define & h_{\mbox{\scriptsize{eff}}}{\Big (}  \tau = \frac{k \tau_p}{M} , \nu = \frac{l \nu_p}{N} {\Big )}.
\end{eqnarray}From the I/O relation in (\ref{eqnio1}) it is clear that 
for detecting the DD domain information symbols from $y_{\mbox{\scriptsize{dd}}}[k,l]$, it suffices to have knowledge of
$h_{\mbox{\scriptsize{eff}}}[k,l]$ only. The receiver does not need to acquire $h_{\mbox{\scriptsize{phy}}}(\tau, \nu)$. Instead it acquires $h_{\mbox{\scriptsize{eff}}}[k,l]$ directly from the channel response to pilots in the discrete DD domain. This makes the Zak-OTFS I/O relation applicable to any model of the underlying physical channel and is therefore \emph{model-free}. Next, we consider the acquisition of $h_{\mbox{\scriptsize{eff}}}[k,l]$.

Consider transmitting a pilot signal $x_{\mbox{\scriptsize{p,dd}}}[k,l]$ together with a data signal $x_{\mbox{\scriptsize{d,dd}}}[k,l]$ within a single Zak-OTFS subframe. Each signal is quasi-periodic, hence is completely specified by the values it takes within the \emph{fundamental region} ${\mathcal D} = \{ (k,l) \, | \, k=0,1,\cdots, M-1, l =0,1,\cdots, N-1\}$.
The pilot signal $x_{\mbox{\scriptsize{p,dd}}}[k,l]$
is determined by a unit energy Dirac-delta impulse at the pilot location $(k_p, l_p) \in {\mathcal D}$ and repeats along the delay and Doppler axis by integer multiples of the delay and Doppler period respectively. It is given by
\begin{eqnarray}
\label{eqnexprxpdd1}
    x_{\mbox{\scriptsize{p,dd}}}[k,l] & \hspace{-3mm} = & \hspace{-3mm} \sum\limits_{n,m \in {\mathbb Z}}  e^{j 2 \pi n \frac{l_p}{N}} \, \delta[k - k_p - nM] \, \delta[l - l_p - mN]. \nonumber \\
\end{eqnarray}The data signal $x_{\mbox{\scriptsize{d,dd}}}[k,l]$ is determined by the unit energy information symbols $x[k_0, l_0]$ (${\mathbb E}\left[ \vert x[k_0, l_0] \vert^2 \right] = 1$) at locations
$(k_0, l_0) \in {\mathcal I}$, where ${\mathcal I} \subset {\mathcal D}$, and is given by
\begin{eqnarray}
    x_{\mbox{\scriptsize{d,dd}}}[k,l] & =  \hspace{-3mm} &  \hspace{-3mm} \sum\limits_{n,m \in {\mathbb Z}} \sum\limits_{(k_0, l_0) \in {\mathcal I}} {\Bigg (} x[k_0,l_0] \, e^{j 2 \pi \frac{l_0}{N}} \, \delta[k - k_0 - nM] \, \nonumber \\
    & & \hspace{35mm} \delta[l - l_0 - mN] {\Bigg )}.
\end{eqnarray}The transmit DD domain signal is
\begin{eqnarray}
    x_{\mbox{\scriptsize{dd}}}[k,l] & = & \sqrt{\frac{E_d}{\vert{\mathcal I}\vert}}  \, x_{\mbox{\scriptsize{d,dd}}}[k,l] \, +  \, \sqrt{E_p} x_{\mbox{\scriptsize{p,dd}}}[k,l].
\end{eqnarray}The data signal has energy $\sum\limits_{k=0}^{M-1} \sum\limits_{l=0}^{N-1} \left\vert  \sqrt{\frac{E_d}{\vert{\mathcal I}\vert}} \, x_{\mbox{\scriptsize{d,dd}}}[k,l] \right\vert^2 = E_d $ and the pilot signal has energy $E_p$. The ratio $E_p/E_d$ is the ratio of pilot power to data power (PDR). 

For simplicity, we first consider channel estimation
in the absence of interference from data, and we let ${\mathcal S}$ denote the support of the effective channel $h_{\mbox{\scriptsize{eff}}}[k,l]$. From (\ref{eqnio1}) and (\ref{eqnexprxpdd1}), the received pilot is given by
\begin{eqnarray}
\label{eqn83ggd70}
    h_{\mbox{\scriptsize{eff}}}[k,l] *_{\sigma} \left( \sqrt{E_p} \, x_{\mbox{\scriptsize{p,dd}}}[k,l] \right) &  \hspace{-3mm} =  &  \hspace{-3mm} \sqrt{E_p} \hspace{-1mm} \sum\limits_{n,m \in {\mathbb Z}} h_{n,m}[k,l].
\end{eqnarray}The $(n,m)$-th term $h_{n,m}[k,l]$ is the channel response to the Dirac-delta impulse of the quasi-periodic pilot signal $x_{\mbox{\scriptsize{p,dd}}}[k,l]$ located at $(k_p + nM, l_p + mN)$, and is given by
\begin{eqnarray}
\label{eqnhnm}
 h_{\mbox{\scriptsize{eff}}}[k,l] *_{\sigma} \, \left( e^{j 2 \pi n \frac{l_p}{N}} \delta[k - k_p -nM] \, \delta[l - l_p - mN] \right) & & \nonumber \\
    & & \hspace{-72mm} = {\Big (} h_{\mbox{\scriptsize{eff}}}[k -k_p - nM,l - l_p - mN]  \, e^{j 2 \pi \frac{n l_p}{N}} \nonumber \\
    & & \hspace{-60mm} \, e^{j 2 \pi \frac{(l - l_p -mN) (k_p + nM) }{MN}} {\Big )}.
\end{eqnarray}The support ${\mathcal S}_{n,m}$ of $h_{n,m}[k,l]$ is ${\mathcal S} + (k_p + nM, l_p + mN)$. 
The \emph{crystallization condition} is ${\mathcal S}_{n,m} \cap {\mathcal S}_{n',m'} = \phi$ for $(n,m) \ne (n',m')$, and when it is satisfied, there is no DD domain aliasing. We have emphasized in \cite{zakotfs2}
that the crystallization condition is satisfied when\footnote{\footnotesize{For any real number $x \in {\mathbb R}$, $\lceil x \rceil$ is the smallest integer greater than or equal to $x$.}}
\begin{eqnarray}
\label{cryscnd}
    k_{max} \Define \left\lceil \frac{M \tau_{max}}{\tau_p} \right\rceil & < & M,  \nonumber \\
    l_{max} \Define \left\lceil \frac{2 N \nu_{max}}{\nu_p} \right\rceil & < & N.
\end{eqnarray}Here $\tau_{max} > 0$ and $\nu_{max} > 0$ are respectively the maximum possible delay and Doppler shift induced by any physical channel path.
The first condition in (\ref{cryscnd}) is that the channel delay spread $\tau_{max}$ is less than the delay period $\tau_p$, and the second condition is that the channel Doppler spread $2 \nu_{max}$ is less than the Doppler period $\nu_p$. We refer the reader to \cite{zakotfs2}, Section II-D for a more extensive discussion of crystallization conditions.

When the crystallization condition holds
\begin{eqnarray}
    h_{0,0}[k,l] & = & h_{\mbox{\scriptsize{eff}}}[k -k_p,l - l_p]  \, e^{j 2 \pi \frac{k_p(l - l_p)}{MN}}
\end{eqnarray}for $(k,l) \in (k_p, l_p) + {\mathcal S}$ and therefore 
\begin{eqnarray}
h_{\mbox{\scriptsize{eff}}}[k,l] & = & h_{0,0}[k+k_p,l+l_p] \, e^{-j 2 \pi \frac{k_p l}{MN}}
\end{eqnarray} for $(k,l) \in {\mathcal S}$. For $(k,l) \in {\mathcal S} + (k_p, l_p)$, the received pilot response (AWGN-free) is simply $h_{0,0}[k,l]$ since the support sets of $h_{n,m}[k,l]$, $n,m \in {\mathbb Z}$ do not overlap when the crystallization condition is satisfied. Hence, the taps of the effective channel filter can simply be read off from the received pilot response within ${\mathcal S} + (k_p, l_p)$. As a result the Zak-OTFS I/O relation in (\ref{eqnio1}) is predictable, i.e., the AWGN-free channel response to any arbitrary input $x_{\mbox{\scriptsize{dd}}}[k,l]$ can be accurately predicted to be $h_{\mbox{\scriptsize{eff}}}[k,l] *_{\sigma} \, x_{\mbox{\scriptsize{dd}}}[k,l]$.

\begin{figure}
     \centering
        \includegraphics[width=5.4cm,height=4.5cm]{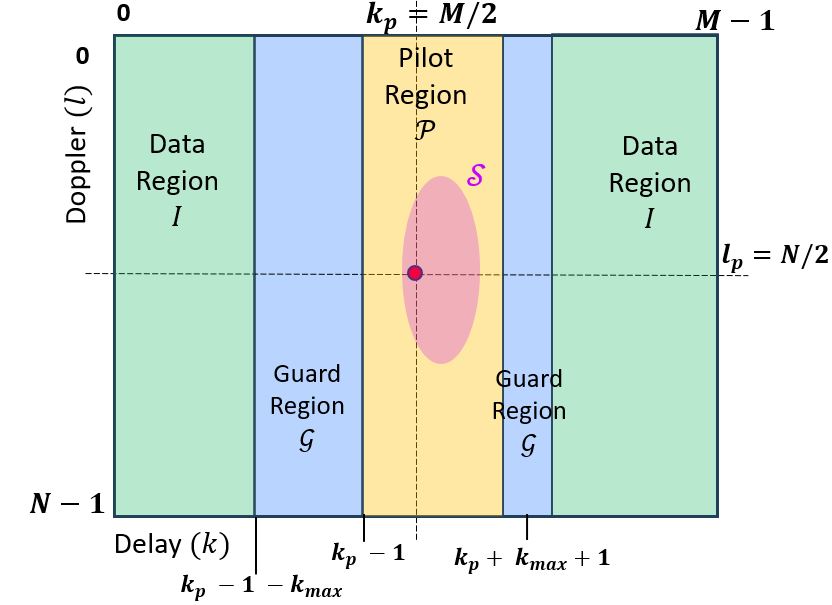}
        \vspace{-2mm}
        \caption{Zak-OTFS DD frame with single pilot (depicted by a red dot) at $(k_p, l_p)$. The pink shaded ellipse depicts the support set of the channel response to the pilot (i.e., ${\mathcal S} \, + \, (k_p, l_p)$).}
        \label{fig_singlepilot}
        \vspace{-3mm}
    \end{figure}

We now consider channel estimation in the presence of interference from data. We transmit a pilot at location $(k_p, l_p)$, and we surround it with pilot and guard regions where no data is transmitted. The pilot region ${\mathcal P}$ is given by
\begin{eqnarray}
{\mathcal P} & = & \{ (k,l) \, | \, k_p - 1 \leq k \leq k_p + k_{max}  \nonumber \\
& & \hspace{10mm} l =0,1,\cdots, N-1 \}.
\end{eqnarray}The guard region ${\mathcal G}$ separates the pilot region from the data region comprising locations $(k_0, l_0) \in {\mathcal I}$. Fig.~\ref{fig_singlepilot} shows pilot, guard and data regions as strips in the Zak-OTFS subframe that run parallel to the Doppler axis.
{We use the pulsones in the pilot region (yellow strip) to acquire the taps of  $h_{\mbox{\scriptsize{eff}}}[k,l]$, and therefore they do not carry information symbols. We do not transmit data in the entirety of the yellow strip so that we are able to acquire the I/O relation for a wide range of Doppler spreads (support of $h_{\mbox{\scriptsize{eff}}}[k,l]$ is shown as a pink ellipse).}

{The guard region on the left of the pilot region is bigger compared to that on the right. This is because, the channel path delays are positive and therefore the delay spread of the effective DD domain channel filter $h_{\mbox{\scriptsize{eff}}}[k,l]$ is asymmetric, i.e., more for positive delay tap values and less for negative delay tap values. The guard region on the left is required to minimize interference to the pilot region from information symbols in the data region to the left and the guard region on the right is required to minimize interference from the data region to the right. Due to the asymmetric delay spread of $h_{\mbox{\scriptsize{eff}}}[k,l]$, the symbols in the data region to the left can spread farther to the right when compared to the symbols in the data region to the right which can spread only a few taps to the left.}

{For the single pilot in Fig.~\ref{fig_singlepilot}, the delay axis width of the guard region to the left of the pilot region is $k_{max}$ taps and that of the guard region to the right of the pilot region is one tap. The width of the pilot region is $(k_{max} + 2)$ taps. Since no information symbols are transmitted in the guard and the pilot regions, the fractional pilot overhead (i.e., ratio of the number of pulsones which do not carry information to the total number of pulsones) is $(2k_{max} + 3)/M$. }
    \begin{figure}
     \centering
        \includegraphics[width=8.2cm,height=7.9cm]{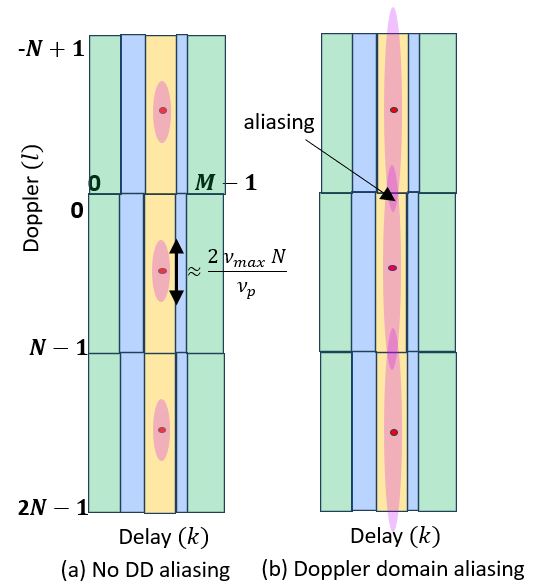}
        \vspace{-2mm}
        \caption{Doppler domain aliasing. Assuming that the channel path Doppler shift lies in $[-\nu_{max} \,,\, \nu_{max}]$, the maximum Doppler spread is $2 \nu_{max}$. The Doppler resolution is $\nu_p/N$ (size of each Doppler bin) and the spread of the channel response is roughly $2 \nu_{max}N/\nu_p$ bins along the Doppler axis. (a) Doppler spread is less than $N$ Doppler axis bins, and the ellipses representing the channel response to constituent pilot impulses do not overlap. (b) Doppler spread is greater than $N$ bins, and the ellipses overlap, preventing accurate estimation of the effective channel $h_{\mbox{\scriptsize{eff}}}[k,l]$.}
        \label{fig_singlepilotaliasing}
        \vspace{-3mm}
    \end{figure}
    
Fig.~\ref{fig_singlepilotaliasing} illustrates the phenomenon of Doppler domain aliasing. In Fig.~$2$(b) the crystallization condition is not satisfied since the
channel Doppler spread $2 \nu_{max}$ is greater than the Doppler period $\nu_p$. In this case Doppler domain aliasing prevents accurate estimation of the effective channel $h_{\mbox{\scriptsize{eff}}}[k,l]$. One solution is to increase $\nu_p$ so that $\nu_p > 2 \nu_{max}$, but this changes the Zak-OTFS delay and Doppler period parameters. In this paper, we design interleaved pilots that resolve Doppler domain aliasing, thus enabling accurate estimation of the effective channel $h_{\mbox{\scriptsize{eff}}}[k,l]$ without changing the period parameters.      


\section{Two Interleaved Pilots}
\label{sectwopilot}
\begin{figure}
     \centering
        \includegraphics[width=7.4cm,height=6.5cm]{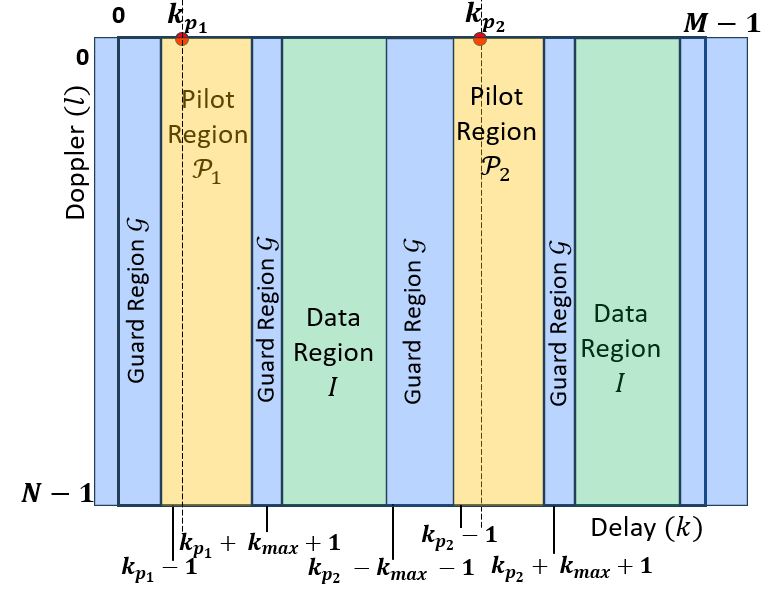}
        \vspace{-2mm}
        \caption{Zak-OTFS subframe with two interleaved pilots (indicated by red dots) at $(k_{p_1}, 0)$ and $(k_{p_2}, 0)$.}
        \label{fig_doublepilot}
        \vspace{-3mm}
    \end{figure}

For simplicity, we suppose $2 \nu_p > 2 \nu_{max} \geq \nu_p$. We transmit two interleaved pilots at locations $(k_{p_1}, 0)$ and $(k_{p_2}, 0)$ in ${\mathcal D}$, and Fig.~\ref{fig_doublepilot} illustrates how we surround each pilot with pilot and guard regions. {The purpose of data, guard and pilots regions is the same as for the single pilot Zak-OTFS frame in Fig.~\ref{fig_singlepilot}. Data is not transmitted in the pilot and guard regions.
Just as in Fig.~\ref{fig_singlepilot}, in Fig.~\ref{fig_doublepilot} also the pilot region (yellow strip) is used to acquire $h_{\mbox{\scriptsize{eff}}}[k,l]$.}

\begin{figure}
     \centering
        \includegraphics[width=7.4cm,height=6.5cm]{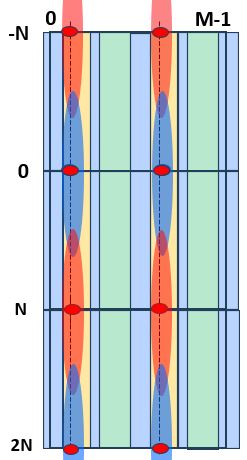}
        \vspace{-2mm}
        \caption{Doppler domain aliasing with two superimposed pilots.}
        \label{fig_doublepilotaliasing}
        \vspace{-3mm}
    \end{figure}

Fig.~\ref{fig_doublepilotaliasing} illustrates the channel responses to the impulses forming the two pilots. For $i=1,2$, the response $h_{0,0}^{(i)}[k,l]$ to the impulse at $(k_{p_i}, 0)$
is shown in blue, and the response $h_{0,1}^{(i)}[k,l]$
to the impulse at $(k_{p_i}, N)$ is shown in red.
Since $2 \nu_p > 2 \nu_{max} \geq \nu_p$, it is only the response of adjacent impulses that overlap along the Doppler axis. 

For simplicity, we consider channel estimation in the
absence of noise, and in the absence of interference from data. The pilot region ${\mathcal P}_1$ is given by
\begin{eqnarray}
        {\mathcal P}_1 & = & \{  (k,l) \, | \, k_{p_1} - 1 \leq k \leq k_{p_1} + k_{max}, \nonumber \\
        & & \hspace{20mm} l =0,1,\cdots, N-1 \}.
    \end{eqnarray}It follows from (\ref{eqn83ggd70}) that the response $y_{\mbox{\scriptsize{dd}}}^{(1)}[k,l]$ to the pilot at $(k_{p_1}, 0)$, received in ${\mathcal P}_1$, is given by\footnote{\footnotesize{{To highlight the main idea, we have made the simplifying assumption that pilot spacing along the delay axis is such that the responses to the two pilots do not overlap. We make no such assumption in Section \ref{simsec}, and in the
    simulations reported there, the pilot responses alias along the delay domain.}}} 
    \begin{eqnarray}
    \label{eqnyddkl}
        y_{\mbox{\scriptsize{dd}}}^{(1)}[k,l] &  = & \sqrt{\frac{E_p}{2}} \, h_{0,0}^{(1)}[k,l] \, + \, \sqrt{\frac{E_p}{2}} \, h_{0,1}^{(1)}[k,l],
    \end{eqnarray}for $(k,l) \in {\mathcal P}_1$.\footnote{\footnotesize{Since there are two interleaved pilots, the energy of each pilot is now $E_p/2$.}}
It now follows from (\ref{eqnhnm}) that for $-1 \leq k \leq k_{max}$ and $l=0,1,\cdots, N-1$
    \begin{eqnarray}
    \label{eqnyddkl}
        y_{\mbox{\scriptsize{dd}}}^{(1)}[k + k_{p_1},l] & \hspace{-3mm} = & \hspace{-3mm} \sqrt{\frac{E_p}{2}}  \, h_{\mbox{\scriptsize{eff}}}[k,l] \, e^{j 2 \pi \frac{l k_{p_1}}{MN} }  \nonumber \\
        & &  \hspace{-2mm}   + \, \sqrt{\frac{E_p}{2}} \, h_{\mbox{\scriptsize{eff}}}[k,l-N] \, e^{j 2 \pi \frac{(l - N) k_{p_1}}{MN} }.
    \end{eqnarray}Note that $y_{\mbox{\scriptsize{dd}}}^{(1)}[k + k_{p_1},l]$ is a linear combination of the unknown taps $h_{\mbox{\scriptsize{eff}}}[k,l]$ and $h_{\mbox{\scriptsize{eff}}}[k,l-N]$. Let $y_{\mbox{\scriptsize{dd}}}^{(2)}[k,l]$ denote the response to the pilot at $(k_{p_2}, 0)$ received in the pilot region ${\mathcal P}_2$. Therefore
        \begin{eqnarray}
    \label{eqnyddkl2}
         y_{\mbox{\scriptsize{dd}}}^{(2)}[k + k_{p_2},l] & \hspace{-3mm} = & \hspace{-3mm} \sqrt{\frac{E_p}{2}} \, h_{\mbox{\scriptsize{eff}}}[k,l] \, e^{j 2 \pi \frac{l k_{p_2}}{MN} }  \nonumber \\
        & & \hspace{-2mm}  + \, \sqrt{\frac{E_p}{2}} \, h_{\mbox{\scriptsize{eff}}}[k,l-N] \, e^{j 2 \pi \frac{(l - N) k_{p_2}}{MN} },
    \end{eqnarray}for $ -1 \leq k \leq k_{max}$, $l=0,1,\cdots, N-1 $. {When $k_{p_1} \ne k_{p_2}$ (mod $M$), equations (\ref{eqnyddkl}) and (\ref{eqnyddkl2}) are linearly independent, and it is possible to solve for $h_{\mbox{\scriptsize{eff}}}[k,l]$ and $h_{\mbox{\scriptsize{eff}}}[k,l-N]$}.\footnote{\footnotesize{{In general, the received samples $y_{\mbox{\scriptsize{dd}}}^{(1)}[k + k_{p_1},l]$ and $y_{\mbox{\scriptsize{dd}}}^{(2)}[k + k_{p_2},l]$ also contain noise and interference from data symbols.
    Therefore, for the simulations reported in Section \ref{simsec}, we consider the least squares (LS) estimate of the channel tap coefficients $h_{\mbox{\scriptsize{eff}}}[k,l]$ and $h_{\mbox{\scriptsize{eff}}}[k,l-N]$. This also holds for the general case of $Q$ interleaved pilots discussed in Section \ref{secmultipilot}, where we need to estimate $Q$ channel tap coefficients from $Q$ linear equations with additive noise and data interference.}} }
    {The least squares (LS) estimate of $(h_{\mbox{\scriptsize{eff}}}[k,l], h_{\mbox{\scriptsize{eff}}}[k,l-N])^T$ is given by (\ref{eqn72443}) (see top of this page).}
    \begin{figure*}
    \vspace{-7mm}
    \begin{eqnarray}
    \label{eqn72443}
    \begin{bmatrix}
        \widehat{h_{\mbox{\scriptsize{eff}}}}[k,l] \\
        \widehat{h_{\mbox{\scriptsize{eff}}}}[k,l-N] 
    \end{bmatrix}
    & = &
    \sqrt{\frac{2}{E_p}}{\Bigg (}\begin{bmatrix}
    e^{j 2 \pi \frac{l k_{p_1}}{MN} } & e^{j 2 \pi \frac{(l - N)k_{p_1}}{MN} } \\
    e^{j 2 \pi \frac{l k_{p_2}}{MN} } & e^{j 2 \pi \frac{(l - N)k_{p_2}}{MN} }
     \end{bmatrix}
     {\Bigg )}^{-1} \,\,
     \begin{bmatrix}
     y_{\mbox{\scriptsize{dd}}}^{(1)}[k + k_{p_1},l] \\
         y_{\mbox{\scriptsize{dd}}}^{(2)}[k + k_{p_2},l] 
     \end{bmatrix}
     \end{eqnarray}
     \vspace{-4mm}
     \begin{eqnarray*}
         \hline
     \end{eqnarray*}
     \end{figure*}
    When the channel Doppler spread satisfies $2 \nu_p > 2 \nu_{max} \geq \nu_p$, the discrete Doppler domain spread is less than $2N$, and therefore the effective channel taps $h_{\mbox{\scriptsize{eff}}}[k,l]$ can be acquired accurately for all $(k,l) \in {\mathcal S}$.\footnote{\footnotesize{$h_{\mbox{\scriptsize{eff}}}[k,l]$, $l=0,1,\cdots, N-1$ gives the taps for Doppler indices $\{ 0,1,\cdots, N-1 \}$ and $h_{\mbox{\scriptsize{eff}}}[k,l-N]$ gives the taps for Doppler indices $\{ -1, -2, \cdots, -N \}$. }} The I/O relation is predictable because it is possible to acquire the effective channel.

    {{For the Zak-OTFS frame with two interleaved pilots in Fig.~\ref{fig_doublepilot}, the fractional pilot overhead (i.e., ratio of the number of pulsones which do not carry information to the total number of pulsones) is double of that for the single pilot Zak-OTFS frame in Fig.~\ref{fig_singlepilot}, i.e., $2(2k_{max} + 3)/M$. }}

    If the data and guard regions were not present in Fig.~\ref{fig_doublepilot}, the minimum possible delay domain pilot spacing $\min_{n_1, n_2 \in {\mathbb Z}} \left\vert k_{p_2} + n_2 M - k_{p_1} - n_1 M \right\vert$ should be chosen as $k_{max} + 2$ in order to avoid delay domain aliasing between the response to pilots which are adjacent along the delay axis. Since pilots are quasi-periodic with delay period $M$, we have
    $M \geq 2(k_{max} + 2)$ and therefore the delay spread $\tau_{max}$ satisfies $\tau_{max} \leq \tau_p/2 \, - \, 2 \tau_p/M$. We define the effective delay period $\tau_{\mbox{\scriptsize{p,eff}}} = \tau_p/2$ and the effective Doppler period $\nu_{\mbox{\scriptsize{p,eff}}} = 2 \nu_p$, noting that the product $\tau_{\mbox{\scriptsize{p,eff}}} \, \nu_{\mbox{\scriptsize{p,eff}}} = 1$ is unchanged. The I/O relation is predictable when the following \emph{effective crystallization condition}
    is satisfied
    \begin{eqnarray}
    \label{effcnd34}
        k_{max} + 2 \leq \frac{M}{2} \,\, \mbox{\small{and}} \,\, \left\lceil \frac{2 \nu_{max} N}{\nu_p} \right\rceil \, < \,  2N. 
    \end{eqnarray}

    Interleaved pilots make it possible to change the aspect ratio of the crystallization condition without changing the fundamental periods, $\tau_p$ and $\nu_p$. Here, we have illustrated the method of interleaved pilots for the case $2 \nu_p > \nu_{max} \geq \nu_p$. {By symmetry, a similar method applies when we have channel scenarios with large delay spread, larger than $\tau_p$. For example, with $2 \tau_p > \tau_{max} \geq \tau_p$, we apply the same method with the only difference being that the pilots are interleaved along the Doppler axis (instead of the delay axis) and are also multiplied with distinct known unit modulus complex scalars. This is required, since the pilot is periodic along the Doppler axis and therefore without any multiplication with complex scalars the corresponding equations (\ref{eqnyddkl}) and (\ref{eqnyddkl2}) will not be linearly independent. Note that there are important wireless markets which have large cells.} 

    \section{Multiple Interleaved Pilots}
    \label{secmultipilot}
     The generalization to $Q$ interleaved pilots is required when
     $(Q-1)\nu_p \leq 2 \nu_{max} < Q \nu_p$. We transmit $Q$
     interleaved pilots at locations $(k_{p_i}, 0) \in {\mathcal D}, i=1,2 \cdots, Q$,
     separated by pilots regions ${\mathcal P}_i$ and guard regions.
     The response $y_{\mbox{\scriptsize{dd}}}^{(i)}[k,l]$ to the pilot $(k_{p_i}, 0)$ received in ${\mathcal P}_i$ is a linear combination
     of $Q$ distinct taps of $h_{\mbox{\scriptsize{eff}}}[k,l]$. The responses $y_{\mbox{\scriptsize{dd}}}^{(i)}[k,l]$, $i=1,2,\cdots, Q$, yield $Q$ linear equations in the $Q$ unknown taps, and since the pilot locations are distinct, these equations are linearly independent. We estimate the taps $h_{\mbox{\scriptsize{eff}}}[k,l]$, where $k=-1, \cdots, k_{max}$ with $l = -\frac{Q}{2}N, \cdots, 0, \cdots, (\frac{Q}{2} - 1)N$ if $Q$ is even, and $l = -\frac{(Q-1)}{2}N, \cdots, 0, \cdots, \frac{(Q-1)}{2}N$ if $Q$ is odd. {With $Q$ interleaved pilots and guard regions on both sides (just as in Fig.~\ref{fig_singlepilot} and Fig.~\ref{fig_doublepilot} for single and two pilots), the fractional pilot overhead is $Q$ times that for the single pilot, i.e., $Q(2k_{max} + 3)/M$.}

     The delay locations of consecutive pilots along the delay axis must be separated by $k_{max} + 2$ bins, in order to prevent interference between the responses of
     adjacent pilots. Hence, the channel filter $h_{\mbox{\scriptsize{eff}}}[k,l]$ can be accurately acquired and the I/O relation is predictable when the following effective crystallization condition is satisfied.
     \begin{eqnarray}
     \label{eqn82648}
         k_{max} + 2 & \leq & \frac{M}{Q} \,\,\, \mbox{\small{and}} \,\,\, \left\lceil \frac{2 \nu_{max} N}{\nu_p} \right\rceil \, < \, QN,
     \end{eqnarray}which implies
     \begin{eqnarray}
         \tau_{max} & < & \frac{\tau_p}{Q} \,\,,\,\, 2\nu_{max} \, < \, Q \nu_p.
     \end{eqnarray}Therefore, the effective delay period is $\tau_p/Q$ and the effective Doppler period is $Q \nu_p$.

    
\section{Auto-ambiguity of Interleaved Pilots}
\label{autoambig}
For simplicity, we again consider channel estimation
in the absence of noise, and in the absence of interference from data. We have shown that it is possible to read off the taps of the effective DD domain channel filter $h_{\mbox{\scriptsize{eff}}}[k,l]$
from the response to an interleaved pilot. While simple and effective, this is not the maximum likelihood (ML) estimate. We have shown (see \cite{zakotfs3} for more details, also \cite{Hanly24}) that the ML estimator is given by the samples of the cross-ambiguity $A_{y, x_p}[k,l]$ between the received DD domain interleaved pilot $y_{\mbox{\scriptsize{dd}}}[k,l]$ and the transmitted DD domain interleaved pilot $x_{\mbox{\scriptsize{p,dd}}}[k,l]$.  
For $(k,l)$ in the support ${\mathcal S}$ of
$h_{\mbox{\scriptsize{eff}}}[k,l]$ 
\begin{eqnarray}
\label{crossambigeqn3}
    A_{y,x_p}[k,l] & \hspace{-3mm} = & \hspace{-3mm} \sum\limits_{k' = 0}^{M-1} \sum\limits_{l' = 0}^{N-1} {\Big (} y_{\scriptsize{dd}}[k',l'] \, x_{p,\scriptsize{dd}}^*[k' - k,l' - l] \, \nonumber \\
    & & \hspace{15mm} e^{-j 2 \pi l \frac{(k' - k)}{MN}} {\Big )}.
\end{eqnarray}
We have shown (\cite{zakotfs3}, Theorem $6$ of Appendix D) that
\begin{eqnarray}
\label{eqn83652}
    A_{y,x_p}[k,l] & \hspace{-3mm} = & \hspace{-3mm} h_{\mbox{\scriptsize{eff}}}[k,l] \, *_{\sigma} \, A_{x_p, x_p}[k,l] 
\end{eqnarray}where
\begin{eqnarray}
\label{axpxpkl}
    A_{x_p, x_p} [k,l] & \hspace{-3mm} = & \hspace{-3mm} \sum\limits_{k' = 0}^{M-1} \sum\limits_{l' = 0}^{N-1} {\Big (} x_{p,\scriptsize{dd}}[k',l'] \, x_{p,\scriptsize{dd}}^*[k' - k,l' - l] \, \nonumber \\
    &  & \,\,\, \hspace{12mm}  e^{-j 2 \pi l \frac{(k' - k)}{MN}} {\Big )} 
\end{eqnarray}is the auto-ambiguity function of the interleaved pilot
\begin{eqnarray}
\label{xpddqkl}
    x_{p,\scriptsize{dd}}[k,l] & \hspace{-3mm} = & \hspace{-3mm} \sqrt{\frac{E_p}{Q}} \sum\limits_{i=1}^Q \sum\limits_{n,m \in {\mathbb Z}} \delta[k - k_{p_i} - nM] \, \delta[l - mN], \nonumber \\
\end{eqnarray}with pilot locations $(k_{p_i}, 0)$, $i=1,2,\cdots, Q$.
\begin{figure*}
\vspace{-7mm} 
\begin{eqnarray}
\label{eqn0937745}
A_{x_p, x_p} [k,l] & \hspace{-3mm} = & \hspace{-3mm} \frac{E_p}{2} \sum\limits_{n,m \in {\mathbb Z}} \delta[k - (k_{p_1} - k_{p_2} - nM) ] \, \delta[l - mN] \, e^{-j 2 \pi \frac{m k_{p_2}}{M} }  \nonumber \\
& & \, + \,  \frac{E_p}{2} \sum\limits_{n,m \in {\mathbb Z}} \delta[k - (k_{p_2} - k_{p_1} - nM) ] \, \delta[l - mN] \, e^{-j 2 \pi \frac{m k_{p_1}}{M} }  \nonumber \\
& & + \frac{E_p}{2} \sum\limits_{n,m \in {\mathbb Z}} \delta[k  - nM ] \, \delta[l - mN] \, \left(  e^{-j 2 \pi \frac{m k_{p_1}}{M} } \, + \, e^{-j 2 \pi \frac{m k_{p_2}}{M} } \right).
\end{eqnarray}
\vspace{-4mm}
\begin{eqnarray*}
    \hline
\end{eqnarray*}
\end{figure*}

We now express the linear estimation method derived in Section \ref{sectwopilot} in terms of ambiguity functions. The auto-ambiguity function for two
interleaved pilots ($Q=2$) is given by (\ref{eqn0937745}) (see top of next page).

When $\vert k_{p_1} - k_{p_2} \vert = M/2$,
it follows from (\ref{eqn0937745}) that the auto-ambiguity function $A_{x_p, x_p} [k,l]$ is non-zero
only on the rectangular lattice $\Lambda_2 = \{ (nM/2, 2mN) \, \vert \, m,n \in {\mathbb Z}\} $. The lattice points are spaced apart by $M/2$ along the delay axis and by $2N$ along the Doppler axis. We translate
$h_{\mbox{\scriptsize{eff}}}[k,l]$ by lattice points
in $\Lambda_2$ to obtain the cross-ambiguity $A_{y, x_p} [k,l]$ in (\ref{eqn83652}). If $2 \nu_{max} < 2 \nu_p$, then the discrete Doppler spread of $h_{\mbox{\scriptsize{eff}}}[k,l]$ does not exceed $2N$, and if $\tau_{max} < \tau_p/2$ then the discrete delay spread of $h_{\mbox{\scriptsize{eff}}}[k,l]$ does not exceed $M/2$. In this case, the translates of the support
${\mathcal S}$ of $h_{\mbox{\scriptsize{eff}}}[k,l]$
by lattice points in $\Lambda_2$ do not overlap.
The crystallization condition (\ref{effcnd34})
is then satisfied and the I/O relation is predictable.
In fact the taps of $h_{\mbox{\scriptsize{eff}}}[k,l]$ can be read off from $A_{y, x_p} [k,l]$ by restricting
to any fundamental domain of $\Lambda_2$.

{From (\ref{eqn0937745}) it also follows that when $\vert k_{p_1} - k_{p_2} \vert \ne M/2$, the
auto-ambiguity function $A_{x_p, x_p}[k,l]$ is non-zero
only at DD points
\begin{eqnarray}
\label{eqn974652}
(k,l) & = & (k_{p_1} - k_{p_2} + nM, mN), \nonumber \\
(k,l) & = & (k_{p_2} - k_{p_1} + nM, mN), \nonumber \\
\mbox{\small{or}} \,\, (k,l) & = & (nM, mN)
\end{eqnarray}for $n,m \in {\mathbb Z}$. The spacing along the Doppler axis is $N$ rather than $2N$, and when $2 \nu_{max} > \nu_p$ it is not possible to accurately estimate $h_{\mbox{\scriptsize{eff}}}[k,l]$.}

{In the Zak-OTFS frame for two interleaved pilots (see Fig.~\ref{fig_doublepilot}), one can choose the two pilot locations such that there is no data region in between enabling the use of a common guard region which can reduce overhead. However, the two pilot locations will then be closer than $M/2$ delay taps due to which the auto-ambiguity function of the interleaved pilot signal will have a period of only $N$ along the Doppler axis and not $2N$ due to which the I/O relation cannot be acquired for $\nu_p < 2\nu_{max} < 2\nu_p$ resulting in poor error rate performance.}

\begin{figure}
     \centering
        \includegraphics[width=9.1cm,height=6.1cm]{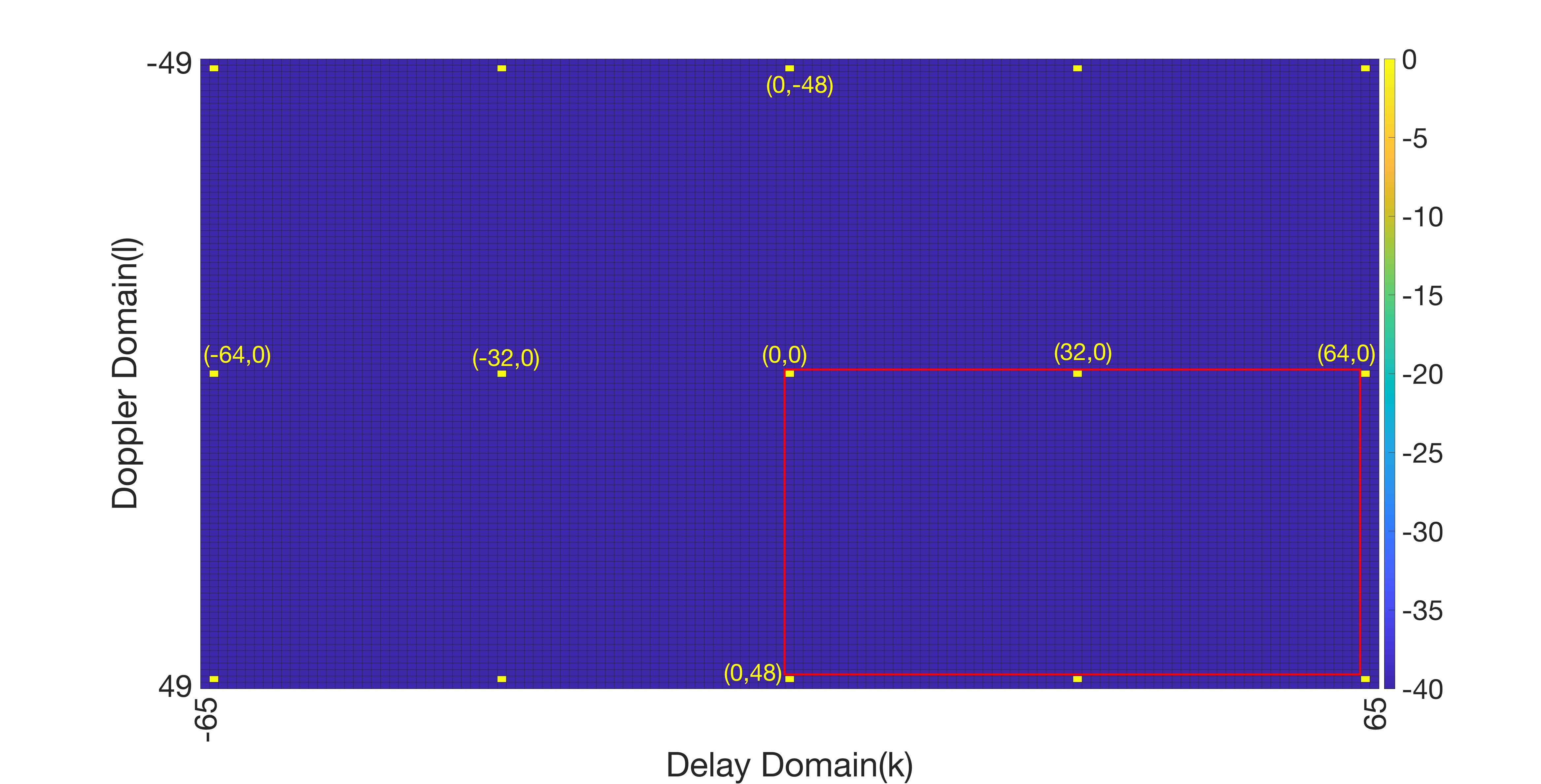}
        \vspace{-2mm}
        \caption{Heatmap showing the magnitude 
        of the auto-ambiguity function $A_{x_p, x_p}[k,l]$ of two interleaved pilots ($Q=2$) at $(0,0)$ and $(32, 0)$ ($M = 64$, $N=24$ ). The pilot spacing is regular ($M/2 = 32$), hence the auto-ambiguity function has period $2N = 48$
        along the Doppler axis.}
        \label{figkp32_autoambig}
        \vspace{-3mm}
    \end{figure}

\begin{figure}
     \centering
        \includegraphics[width=9.4cm,height=6.5cm]{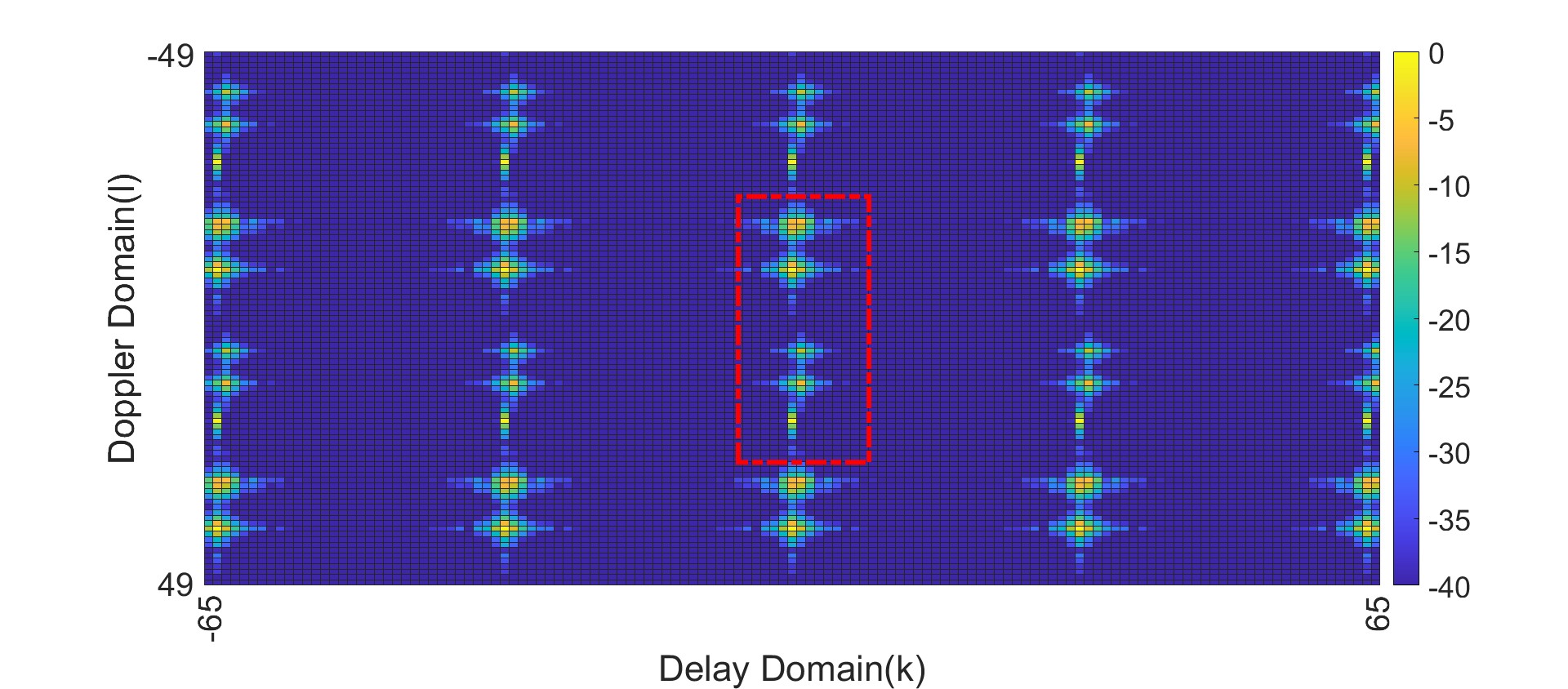}
        \vspace{-2mm}
        \caption{Heatmap showing the magnitude of the cross-ambiguity function $A_{y, x_p}[k,l]$ for two interleaved pilots ($Q=2$) at $(0,0)$ and $(32, 0)$ ($M = 64$, $N=24$ ). Six path Veh-A channel model \cite{EVAITU} with maximum Doppler shift $\nu_{max} = 6$ KHz and $\nu_p = 7.5$ KHz. Two interleaved pilots separated by $M/2 = 32$ along delay axis. The Doppler spread $2 \nu_{max} = 12$ KHz satisfies $\nu_p < 2 \nu_{max} \leq 2 \nu_p$. An accurate estimate for the effective channel $h_{\mbox{\scriptsize{eff}}}[k,l]$ can be read off from samples of the cross-ambiguity function within the support ${\mathcal S}$ of the effective channel (rectangle with red boundary).}
        \label{figkp32_crossambig}
        \vspace{-3mm}
    \end{figure}

\begin{figure}
     \centering
        \includegraphics[width=9.1cm,height=6.1cm]{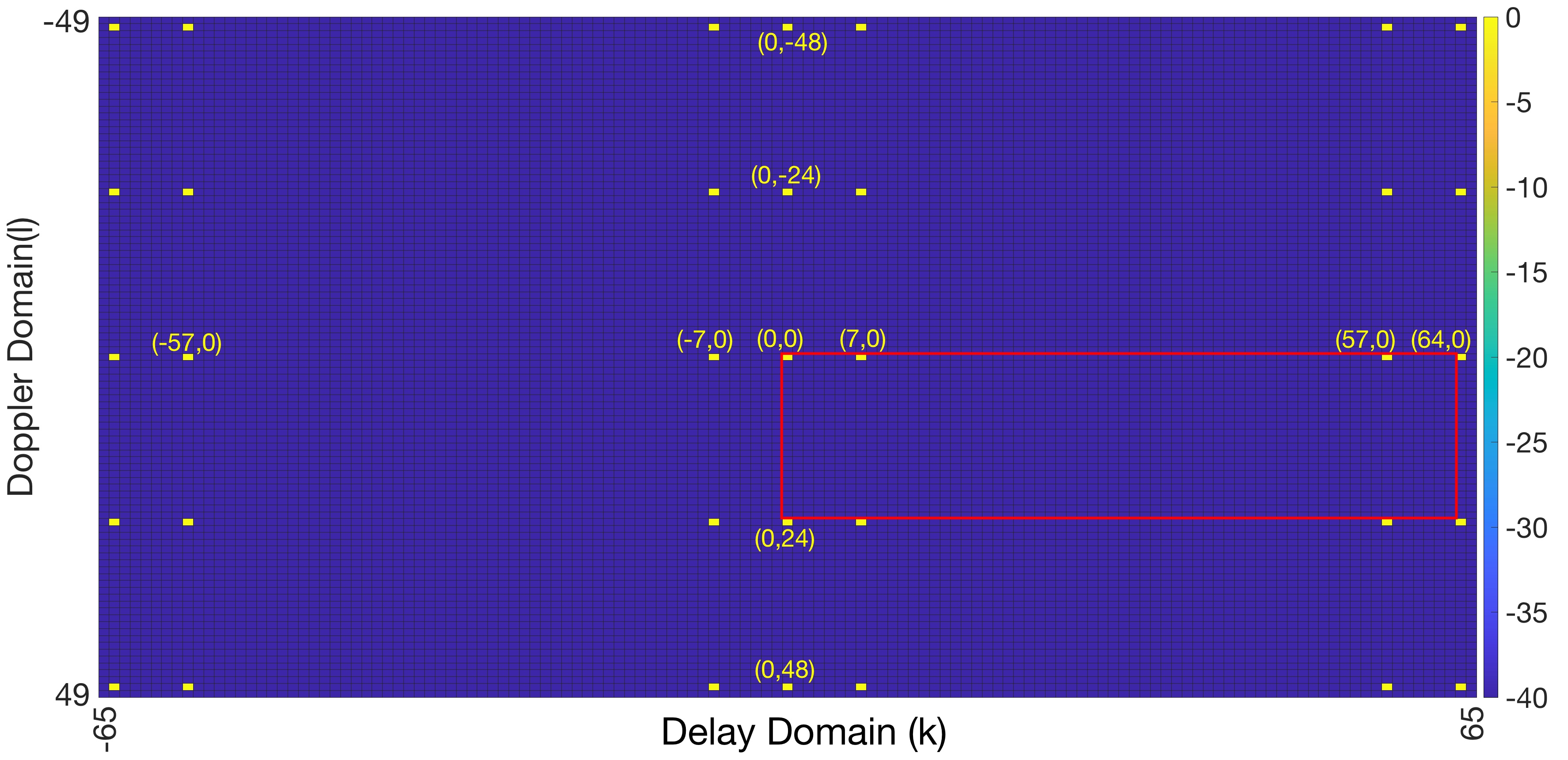}
        \vspace{-2mm}
        \caption{Heatmap showing the magnitude of the auto-ambiguity function $A_{x_p, x_p}[k,l]$ when the pilot signal consists of two interleaved pilots ($Q=2$) at $(0,0)$ and $(7, 0)$. $M = 64$, $N=24$ and the pilot spacing along the delay axis is not equal to $M/2$. The period of the auto-ambiguity function along the Doppler axis is $N$ rather than $2N$ (see (\ref{eqn974652}) for the precise non-zero locations of the auto-ambiguity function).}
        \label{figkp7_autoambig}
        \vspace{-3mm}
    \end{figure}

\begin{figure}
     \centering
        \includegraphics[width=9.4cm,height=6.5cm]{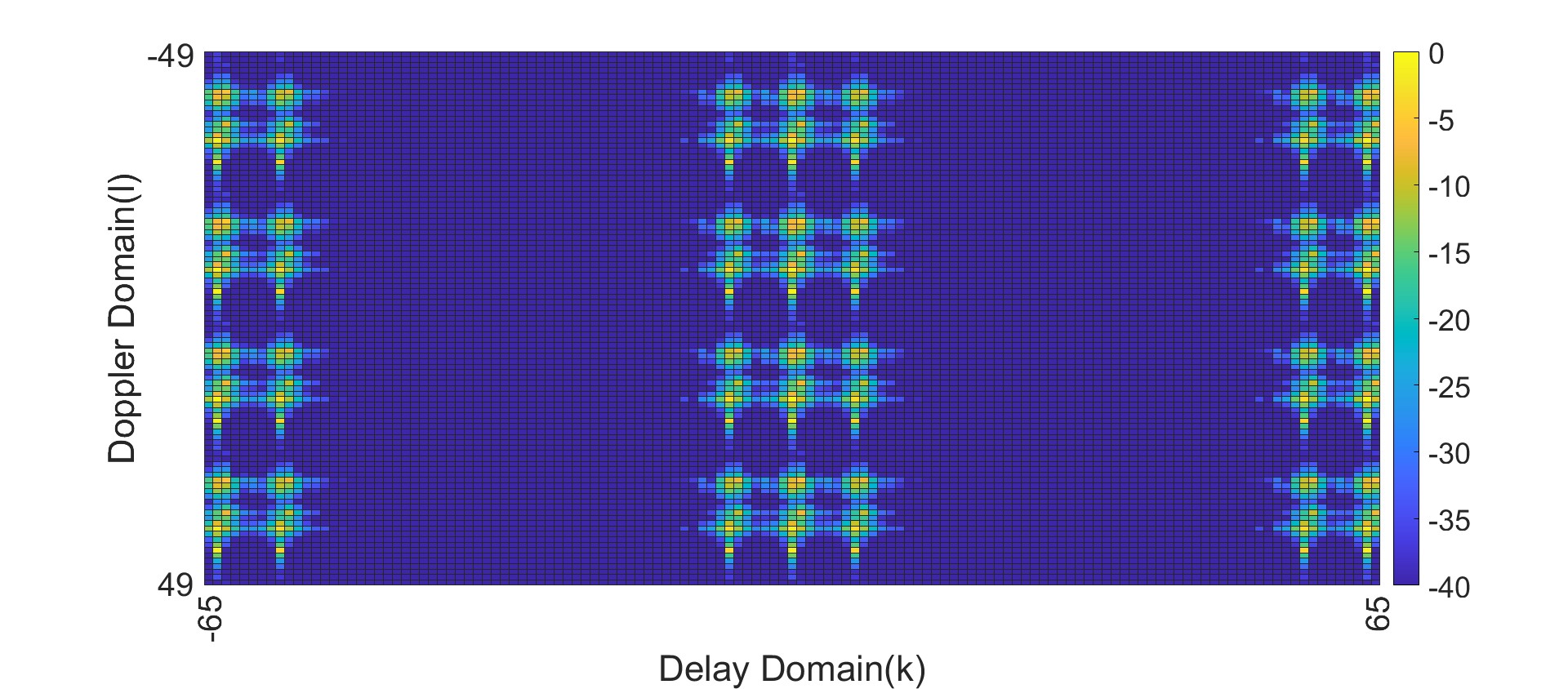}
        \vspace{-2mm}
        \caption{Heatmap showing the magnitude of the cross-ambiguity function $A_{y, x_p}[k,l]$ for two interleaved pilots ($Q=2$) at $(0,0)$ and $(7, 0)$ with $M = 64$, $N=24$. Six path Veh-A channel model \cite{EVAITU} with maximum Doppler shift $\nu_{max} = 6$ KHz and $\nu_p = 7.5$ KHz. Note that the Doppler spread $2 \nu_{max} = 12$ KHz satisfies $\nu_p < 2 \nu_{max} \leq 2 \nu_p$. The pilot spacing along the delay axis is not equal to $M/2$ and the period of the auto-ambiguity function along the Doppler axis is $N$ rather than $2N$. Aliasing of the responses $h_{0,0}^{(i)}[k,l]$ and $h_{0,1}^{(i)}[k,l]$ prevents accurate acquisition of the effective channel $h_{\mbox{\scriptsize{eff}}}[k,l]$ (compare this heatmap with the heatmap in Fig.~\ref{figkp32_crossambig} where there is no aliasing).}
        \label{figkp7_crossambig}
        \vspace{-3mm}
    \end{figure}

{We explain this in detail with the help of Figs. \ref{figkp32_autoambig}, \ref{figkp32_crossambig}, \ref{figkp7_autoambig} and \ref{figkp7_crossambig}.
We consider $M = 64, N= 24$. Fig.~\ref{figkp32_autoambig} shows the heatmap of the auto-ambiguity function of the transmitted two interleaved pilots $(Q=2)$ when the delay domain spacing $\vert k_{p_2} - k_{p_1} \vert = M/2 = 32$. It is observed that the auto-ambiguity function is supported on a rectangular lattice with periods $2N$ and $M/2$ respectively along the Doppler and delay axis as discussed above (based on (\ref{eqn0937745})). However, when the delay domain spacing is only $7$ delay taps (as in Fig.~\ref{figkp7_autoambig}, the auto-ambiguity function is no more supported on a lattice and repeats with a period of $N$ (and not $2N$) along the Doppler axis.}

{In Fig.~\ref{figkp32_crossambig}
and Fig.~\ref{figkp7_crossambig}, we plot the heatmap of the cross-ambiguity between the received and the transmitted two interleaved pilots, when their spacing is $32$ and $7$ respectively. We consider a six-path Veh-A channel \cite{EVAITU} with $\nu_{max} = 6$ KHz and $\nu_p = 7.5$ KHz. Note that the Doppler spread is $2 \nu_{max} = 12$ KHz which is more than $\nu_p$ but less than $2 \nu_p$ and this is why the taps of $h_{\mbox{\scriptsize{eff}}}[k,l]$ can be read-off accurately from the received samples within the red rectangle shown in Fig.~\ref{figkp32_crossambig}, when the delay spacing between the pilots is $M/2 = 32$ taps. However, when the pilot spacing is $7 \ne M/2$, the cross-ambiguity in Fig.~\ref{figkp7_crossambig} looks completely different due to aliasing of responses $h_{0,0}^{(i)}[k,l]$ and $h_{0,1}^{(i)}[k,l]$ due to which accurate acquisition of the effective channel $h_{\mbox{\scriptsize{eff}}}[k,l]$ is not possible.}

We see from (\ref{crossambigeqn3}) that the complexity of ML estimation using the cross-ambiguity
$A_{y, x_p}[k,l]$ is $O(M^2 N^2)$. This compares
unfavorably with the $O(MN)$ complexity of solving the linear system proposed in Sections \ref{sectwopilot} and \ref{secmultipilot}.
Although linear estimation is sub-optimal, numerical simulations in Section \ref{simsec} show that it achieves BER close to that achieved with cross-ambiguity based estimation when the effective crystallization condition is satisfied.

{Next, we analyze how estimation accuracy
of the proposed linear estimation in Section \ref{sectwopilot} and
Section \ref{secmultipilot} depends on the spacing between the interleaved pilots. For simplicity,
we take $Q=2$ and apply linear estimation.
The determinant $\Delta$ of the $2 \times 2$ linear system satisfies
$\vert \Delta \vert = \frac{E_p}{2} \left\vert e^{-j 2 \pi \frac{k_{p_1}}{M}} \, - \, e^{-j 2 \pi \frac{k_{p_2}}{M}} \right\vert $. As the minimum pilot
spacing $\min\left(\left\vert k_{p_1} - k_{p_2} \right\vert \,,\, M - \left\vert k_{p_1} - k_{p_2} \right\vert \right)$
decreases, the determinant $\Delta$ approaches $0$, and the $2 \times 2$ system becomes highly ill-conditioned. We expect estimation accuracy
to degrade as the minimum pilot spacing decreases.}

{The number of interleaved pilots that a user needs for reliable communication depends on the Doppler spread of its channel.
In general, if the Doppler spread $2 \nu_{max}$ lies between $(Q-1)\nu_p$ and $Q\nu_p$, then $Q$ interleaved pilots are required (see discussion in Section \ref{secmultipilot}). With $Q$ interleaved pilots regularly spaced apart by $M/Q$ delay axis taps, the auto-ambiguity function of the transmitted $Q$ interleaved pilots is supported on a DD domain lattice having delay and Doppler period $M/Q$ and $NQ$ respectively. Since the lattice period along the Doppler axis is now $Q$ times greater than the regular lattice period $N$ (with a single pilot), Doppler domain aliasing does not happen as long as the Doppler spread is less than $Q \nu_p$. Therefore, each user will use a different number of interleaved pilots depending on the maximum Doppler spread it experiences. The above result is shown in the following theorem.
}
{
\begin{theorem}
    \label{thm1}
    With $Q$ interleaved pilots and $k_{p_i} = (i-1)M/Q$, $i=1,2,\cdots, Q$, the auto-ambiguity function $A_{x_p,x_p}[k,l]$ in (\ref{eqn83652}) is supported on the DD domain lattice $\Lambda_Q$ given by
    \begin{eqnarray}
        \Lambda_Q & \hspace{-3mm} = & \hspace{-3mm}  {\Big \{} (k,l) \, {\Big \vert} \, k = n\frac{M}{Q}, l = mQN, \, (n,m) \in {\mathbb Z} {\Big \}}.
    \end{eqnarray}
\end{theorem}
\begin{IEEEproof}
See Appendix \ref{prfthm1}.
\end{IEEEproof}
}

\begin{table}[]
            \centering
            \caption{Power-delay profile of Veh-A channel model}
            \begin{tabular}{|c|c|c|c|c|c|c|}
                \hline
                Path number ($i$)       &  1 &  2  &  3  &  4  &  5  &  6  \\
                \hline
                $\tau_i$ ($\mu$s) &  0 & 0.31& 0.71& 1.09 & 1.73 & 2.51 \\
                \hline
                Relative power ($p_i$) dB   &  0 & -1  & -9  & -10 & -15 & -20\\
                \hline
            \end{tabular}
            \label{tab_veha}
            \vspace{2mm}
            
            \vspace{-4mm}
        \end{table}
\section{Numerical simulations}
\label{simsec}

We report simulation results for the Veh-A channel model \cite{EVAITU} which consists of six channel paths. The channel gains $h_i$ are modeled as independent zero-mean complex circularly symmetric Gaussian random variables, normalized so that
$\sum\limits_{i=1}^6 {\mathbb E} \left[ \vert h_i \vert^2 \right] = 1$. Table \ref{tab_veha} lists the power-delay profile for the
six channel paths. The Doppler shift of the $i$-th
path is modeled as $\nu_i = \nu_{max} \, \cos\left( \theta_i \right)$, where $\nu_{max}$ is the maximum Doppler shift of any path, and the variables
$\theta_i, i=1,2,\cdots, 6$, are independent and
uniformly distributed in $[-\pi \,,\, \pi)$.\footnote{\footnotesize{{We consider fractional delay and Doppler shifts, which is representative of real propagation environments. Note that path delays in Table-\ref{tab_veha} are non-integer multiples of the delay domain resolution $1/B$. The Doppler shifts $\nu_i = \nu_{max} \cos(\theta_i)$ are also non-integer multiples of the Doppler domain resolution $1/T$ since $\cos(\theta_i)$ is continuous valued. Also, the proposed scheme of using interleaved pilots for acquisition of the I/O relation is still applicable if we have two paths with same delay but different Doppler shifts, since the required number of interleaved pilots depends only on the maximum spread of $h_{\mbox{\scriptsize{eff}}}[k,l]$ along the Doppler axis.}}}

We consider Zak-OTFS modulation with Doppler spread
$\nu_p = 7.5$ KHz, delay period $\tau_p = 1/\nu_p = 133.33 \, \mu s$, $M=64$, $N=24$. The channel bandwidth
$B = M \nu_p = 0.48$ MHz and the subframe duration $T = N \tau_p = 3.2$ ms.
The information lattice/grid $\Lambda_{\mbox{\scriptsize{dd}}} = \{ (k/B, l/T)$, $k,l \in {\mathbb Z} \, \}$.

{The pulse shaping filter $w_{tx}(\tau, \nu)$ at the transmitter is a factorizable root raised cosine (RRC) filter given by
\begin{eqnarray}
    w_{tx}(\tau, \nu)  =  \sqrt{BT} \, rrc_{\beta_{\tau}}(B \tau) \, rrc_{\beta_{\nu}}(\nu T),
\end{eqnarray}where $0 \leq \beta_{\tau}, \beta_{\nu} \leq 1$ and $rrc_{_{\beta}}(x)$ is the RRC waveform (see \cite{JGProakis}) given by
\begin{eqnarray}
    rrc_{_{\beta}}(x)  & \hspace{-3mm} =  &  \hspace{-3mm} \frac{\sin(\pi x (1 - \beta)) + 4 \beta x \cos(\pi x (1 + \beta))}{\pi x \left( 1 - (4 \beta x)^2 \right)}.
\end{eqnarray} We employ the matched filter $w_{rx}(\tau, \nu) = w_{tx}^*(-\tau, -\nu) \, e^{j 2 \pi \nu \tau}$ at the receiver. We choose roll-off factors $\beta_{\nu} = \beta_{\tau} = 0.6$ so that the effective bandwidth of the Zak-OTFS subframe is
$B' = (1 + \beta_{\tau})B$ and the effective time duration is $T' = (1 + \beta_{\nu})T$. The dependence of the effective time/bandwidth and other characteristics of an OTFS modulated signal on the pulse shaping filter is discussed in detail in \cite{zakotfs4} (see equations $(23)$ and $(26)$ in \cite{zakotfs4} for the expressions of the TD and FD realizations of the Zak-OTFS pulsone).}

We employ MMSE equalization of the matrix-vector form of the Zak-OTFS I/O relation to detect information symbols at the receiver (for more details, see \cite{zakotfs2}).
        \begin{figure}
     \centering
        \includegraphics[width=9.4cm,height=6.5cm]{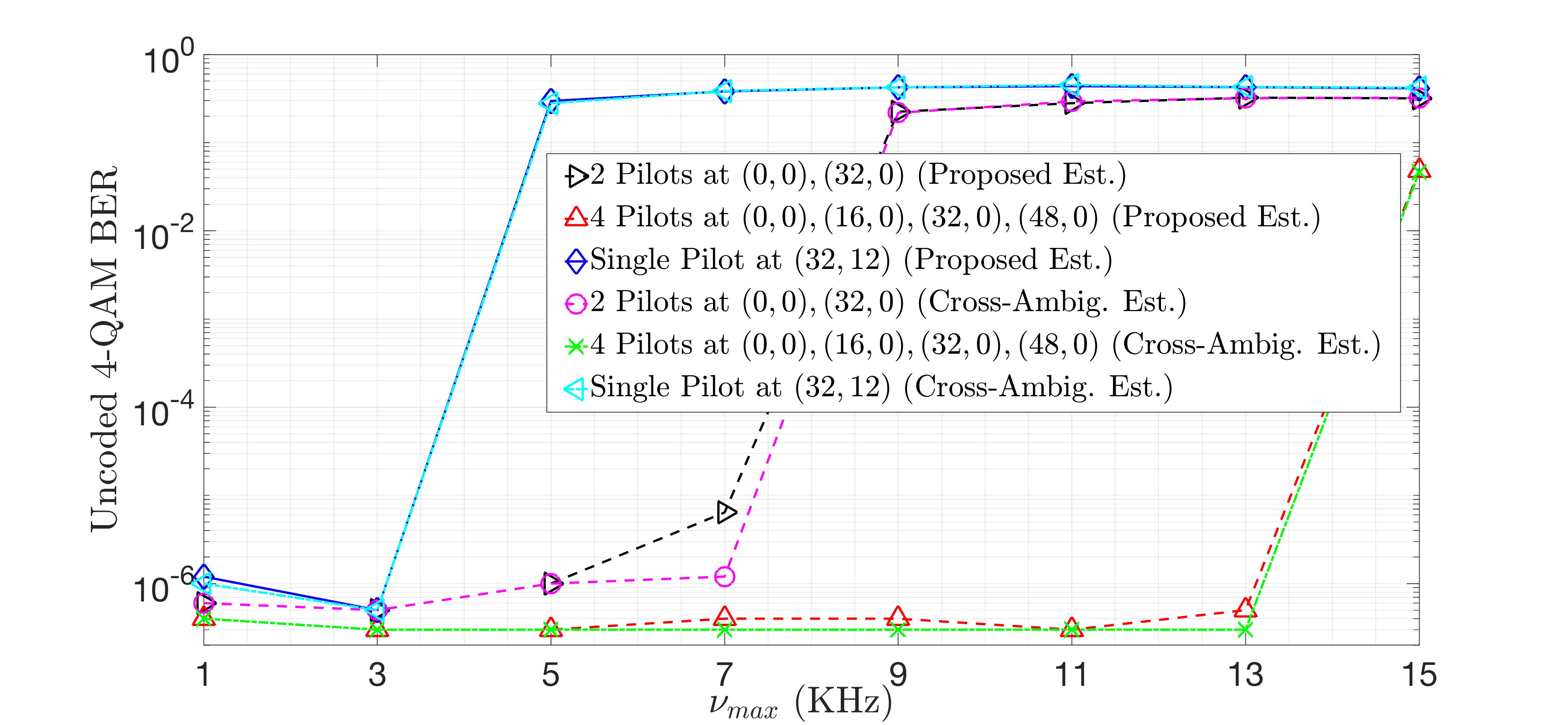}
        \vspace{-2mm}
        \caption{Uncoded $4$-QAM BER as a function of increasing $\nu_{max}$. One, two and four interleaved pilots spaced apart regularly.}
        \label{fig_bervsnumaxtwopilots}
        \vspace{-3mm}
    \end{figure}
    
Fig.~\ref{fig_bervsnumaxtwopilots} plots bit error rate (BER) of uncoded $4$-QAM as a function of increasing $\nu_{max}$. The ratio of pilot to data power (PDR) is $5$ dB and the ratio of received signal power to noise power (data SNR) is $25$ dB. Discrete delay spread $k_{max} = 2$.
{We plot the BER with channel estimates acquired using the proposed linear estimation and that acquired by sampling the cross-ambiguity
function $A_{y,x_p}[k,l]$ at DD taps in the support set of $h_{\mbox{\scriptsize{eff}}}[k,l]$.}

BER performance for a single pilot (cyan and blue curves) degrade sharply for $\nu_{max} > 3.5$ KHz. This is a consequence of
Doppler domain aliasing as $\nu_{max}$ approaches
$\nu_p/2 = 3.75$ KHz (see (\ref{cryscnd}) and the discussion in Section \ref{secsysmodel}).
By contrast, BER performance for two interleaved pilots located at $(0,0)$ and $(M/2, 0)= (32, 0)$ is excellent, even for a Doppler spread $2 \nu_{max} = 14$ KHz which is greater than $\nu_p = 7.5$ KHz but less than $2 \nu_p = 15$ KHz. The simulation is consistent with our theoretical demonstration that two interleaved
pilots can enable reliable communication when the Doppler spread $2 \nu_{max} $ satisfies $\nu_p \leq 2 \nu_{max} < 2 \nu_p$. 

{Fig.~\ref{fig_bervsnumaxtwopilots} also illustrates BER performance for $Q = 4$ interleaved pilots spaced apart regularly along the delay axis (green and red curves). As expected, BER is good for Doppler spreads at most $4 \nu_p$, i.e., $\nu_{max} < 15$ KHz (see (\ref{eqn82648})).}
{Fig.~\ref{fig_bervsnumaxtwopilots}
also shows that the BER performance achieved with the proposed linear estimation method is almost the same as that achieved with the more complex cross-ambiguity based estimation.}

        \begin{figure}
     \centering
        \includegraphics[width=9.6cm,height=6.5cm]{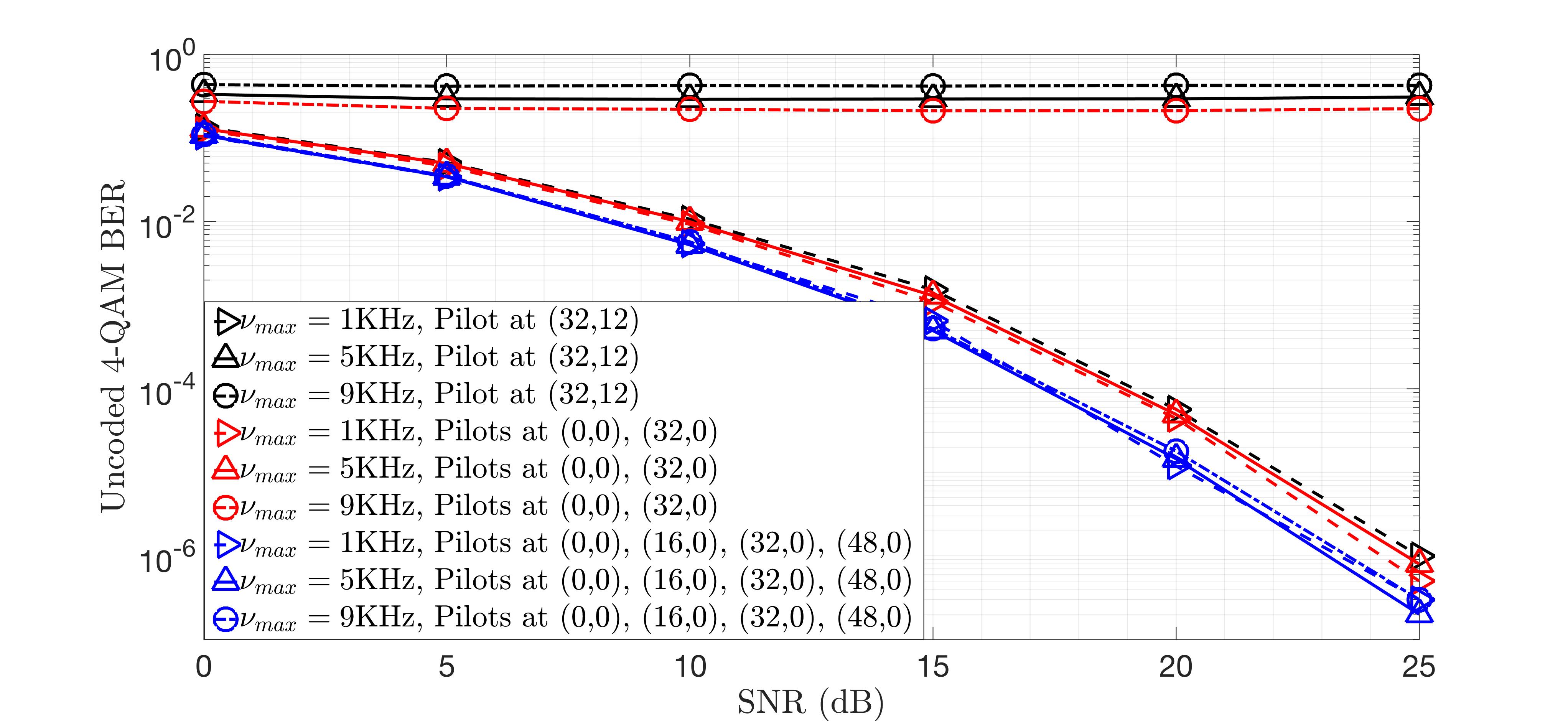}
        \vspace{-2mm}
        \caption{{Uncoded $4$-QAM BER as a function of increasing $SNR$, for $\nu_{max} = 1,5,9$ KHz and fixed PDR$=5$ dB. }}
        \label{fig_bervssnr}
        \vspace{-3mm}
    \end{figure}
{In Fig.~\ref{fig_bervssnr} we plot the uncoded $4$-QAM BER vs. SNR for a fixed $\nu_{max}$ and fixed PDR of $5$ dB. We observe that the conclusions made from Fig.~\ref{fig_bervsnumaxtwopilots} (for SNR$=25$ dB) are in fact valid over a much wider range of SNR values. Note that the BER curve floors for the single pilot case when $\nu_{max} = 5,9$ KHz and it floors for the double pilot case when $\nu_{max} = 9$ KHz. With four interleaved pilots, the BER curve does not floor for $\nu_{max} = 1,5,9$ KHz, since the flooring would only happen when the Doppler spread $2 \nu_{max}$ exceeds $4 \nu_p = 30$ KHz, i.e., when $\nu_{max} > 15$ KHz.}

        \begin{figure}
     \centering
        \includegraphics[width=9.6cm,height=6.5cm]{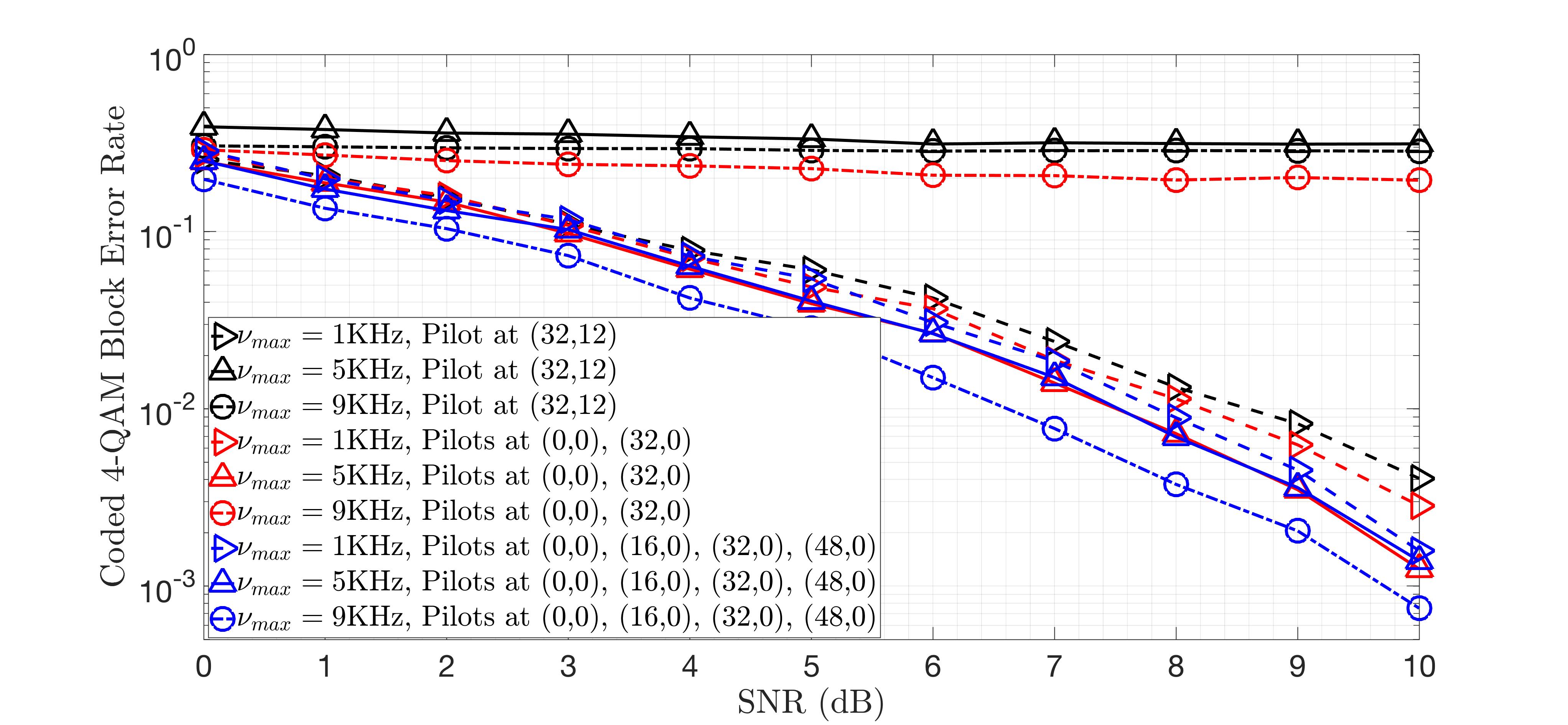}
        \vspace{-2mm}
        \caption{{Coded $4$-QAM block error rate (BLER) as a function of increasing $SNR$, for $\nu_{max} = 1,5,9$ KHz and fixed PDR$=5$ dB. }}
        \label{fig_codedbervssnr}
        \vspace{-3mm}
    \end{figure}
{In Fig.~\ref{fig_codedbervssnr} we plot the coded block error rate (BLER) performance for Zak-OTFS modulation with the proposed interleaved pilots for acquisition of I/O relation. We use the 3GPP 5G NR Low-Density Parity Check (LDPC) codes with code-rate $1/2$ and $4-QAM$ modulation symbols. There is no coding across different Zak-OTFS frames, and therefore the code block size (number of information bits)
for one, two and four interleaved pilots is $1368$, $1200$ and $868$ respectively (larger number of pilots result in a larger pilot and guard region overhead and therefore lesser number of information symbols). From the error plots, we observe error flooring for single pilot with Doppler spread more than $\nu_p = 7.5$ KHz, i.e., $\nu_{max} > \nu_p/2$ ($\nu_{max} = 5, 9$ in the figure), and also for two interleaved pilots with Doppler spread more than $2 \nu_p = 15$ KHz, i.e., $\nu_{max} > 7.5$ KHz ($\nu_{max} = 9$ KHz in the figure). These are exactly the same scenarios where error flooring occurs in the uncoded BER plot of Fig.~\ref{fig_bervssnr}. This is because of primary reason being the same, i.e., the channel in these scenarios does not satisfy the crystallization condition w.r.t. the lattice on which the auto-ambiguity function of the interleaved pilots is supported.}

        \begin{figure}
     \centering
        \includegraphics[width=9.4cm,height=6.5cm]{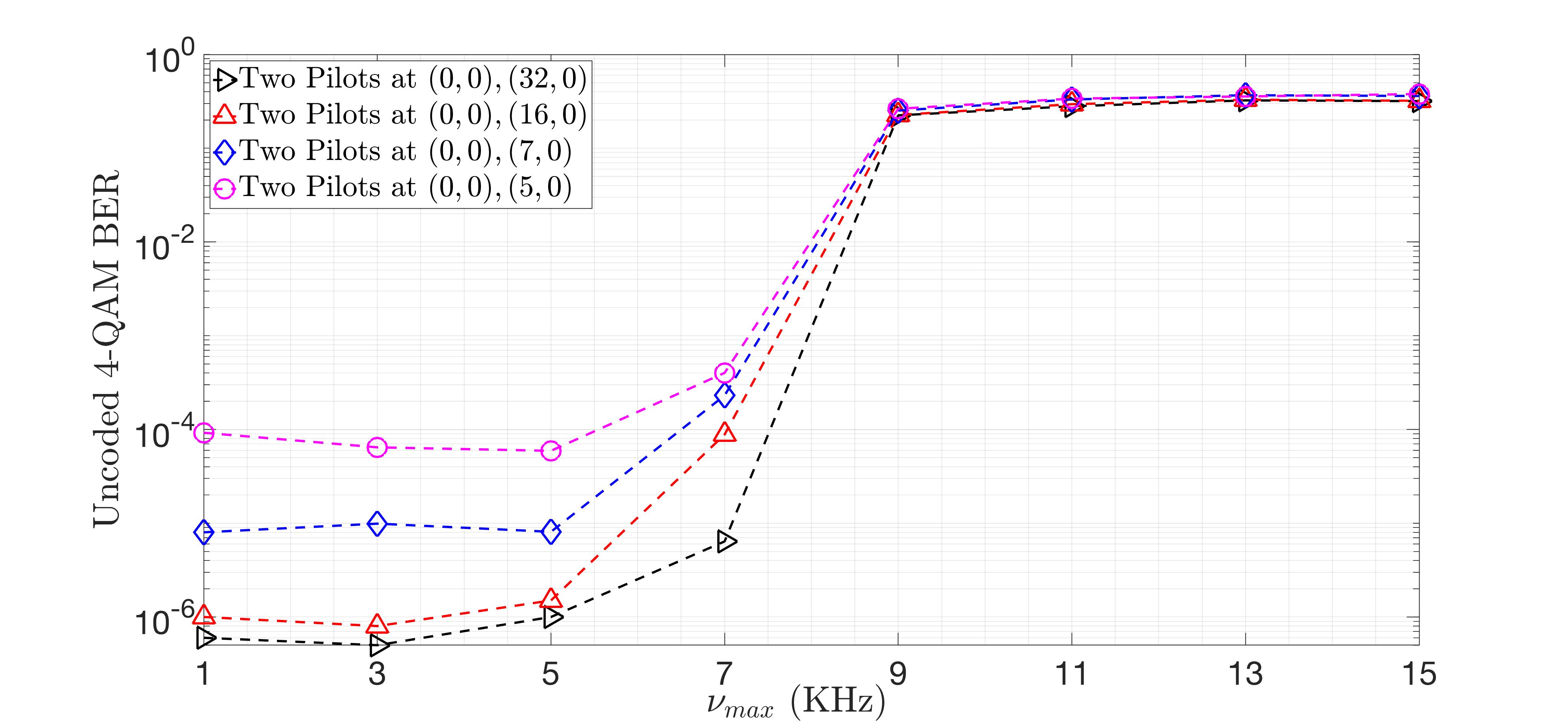}
        \vspace{-2mm}
        \caption{Uncoded $4$-QAM BER as a function of increasing $\nu_{max}$. Two interleaved pilots with irregular spacing.}
        \label{fig_bervsnumaxtwopilotsirr}
        \vspace{-3mm}
    \end{figure}Fig.~\ref{fig_bervsnumaxtwopilotsirr} plots the BER performance for different spacing between two interleaved pilots. Proposed linear estimation of the taps of $h_{\mbox{\scriptsize{eff}}}[k,l]$ is considered. BER performance degrades as pilot spacing decreases, consistent with our discussion in Section \ref{autoambig}.

        \begin{figure}
     \centering
        \includegraphics[width=9.1cm,height=6.1cm]{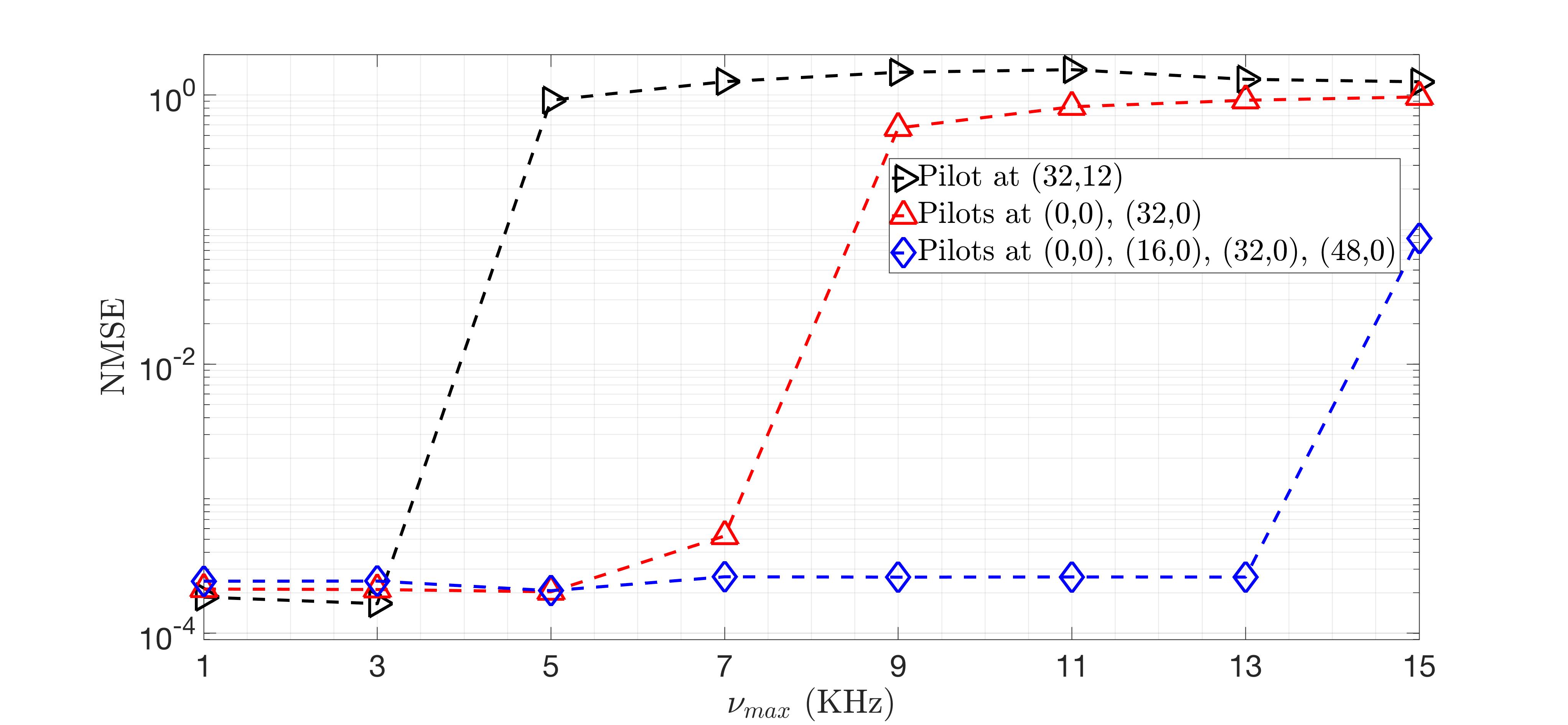}
        \vspace{-2mm}
        \caption{{Normalized mean square error (NMSE) of the acquired channel estimates using the proposed method, as a function of increasing $\nu_{max}$ for the same setting as in Fig.~\ref{fig_bervsnumaxtwopilots}.}}
        \label{fig_nmse_numax}
        \vspace{-3mm}
    \end{figure}
{In Fig.~\ref{fig_nmse_numax} we plot the normalized mean square error (NMSE) of the proposed interleaved pilots based estimation method as a function of increasing $\nu_{max}$. Let $\widehat{h_{\mbox{\scriptsize{eff}}}}[k,l]$ denote the proposed estimate of ${h_{\mbox{\scriptsize{eff}}}}[k,l]$, then the NMSE is given by ${\mathbb E}\left[ \frac{\sum\limits_{(k,l) \in {\mathbb S}}  \vert {h_{\mbox{\scriptsize{eff}}}}[k,l] -  \widehat{h_{\mbox{\scriptsize{eff}}}}[k,l] \vert^2}{\sum\limits_{(k,l) \in {\mathbb S}} \vert {h_{\mbox{\scriptsize{eff}}}}[k,l] \vert^2} \right] $ where ${\mathcal S}$ is the support set for $h_{\mbox{\scriptsize{eff}}}[k,l]$ and the expectation is w.r.t. the random channel realizations. It is observed that with a single pilot, the estimation accuracy is poor as soon as the Doppler spread exceeds $\nu_p$ i.e., $\nu_{max} > \nu_p/2 = 3.75$ KHz. With two interleaved, the estimation accuracy is poor when the Doppler spread exceeds $2\nu_p$ and with four pilots the accuracy suffers only when the Doppler spread exceeds $4 \nu_p$. This is due to the fact that with $Q$ interleaved pilots (regularly spaced along delay axis), the Doppler period of the auto-ambiguity function of the interleaved pilots is $Q \nu_p$ i.e., $Q$ times more than that for a single pilot and which is why the the cross-ambiguity between the received and the transmitted $Q$ interleaved pilot exhibits Doppler domain aliasing only when $2 \nu_{max} > Q \nu_p$. In Figs.~\ref{fig_bervssnr} and \ref{fig_codedbervssnr} the error rate performance floors only
for those scenarios where $2 \nu_{max} > Q \nu_p$. This confirms that the degradation in the error rate performance is primarily due to the
inaccurate channel estimates (i.e., high NMSE).}

        \begin{figure}
     \centering
        \includegraphics[width=9.1cm,height=6.1cm]{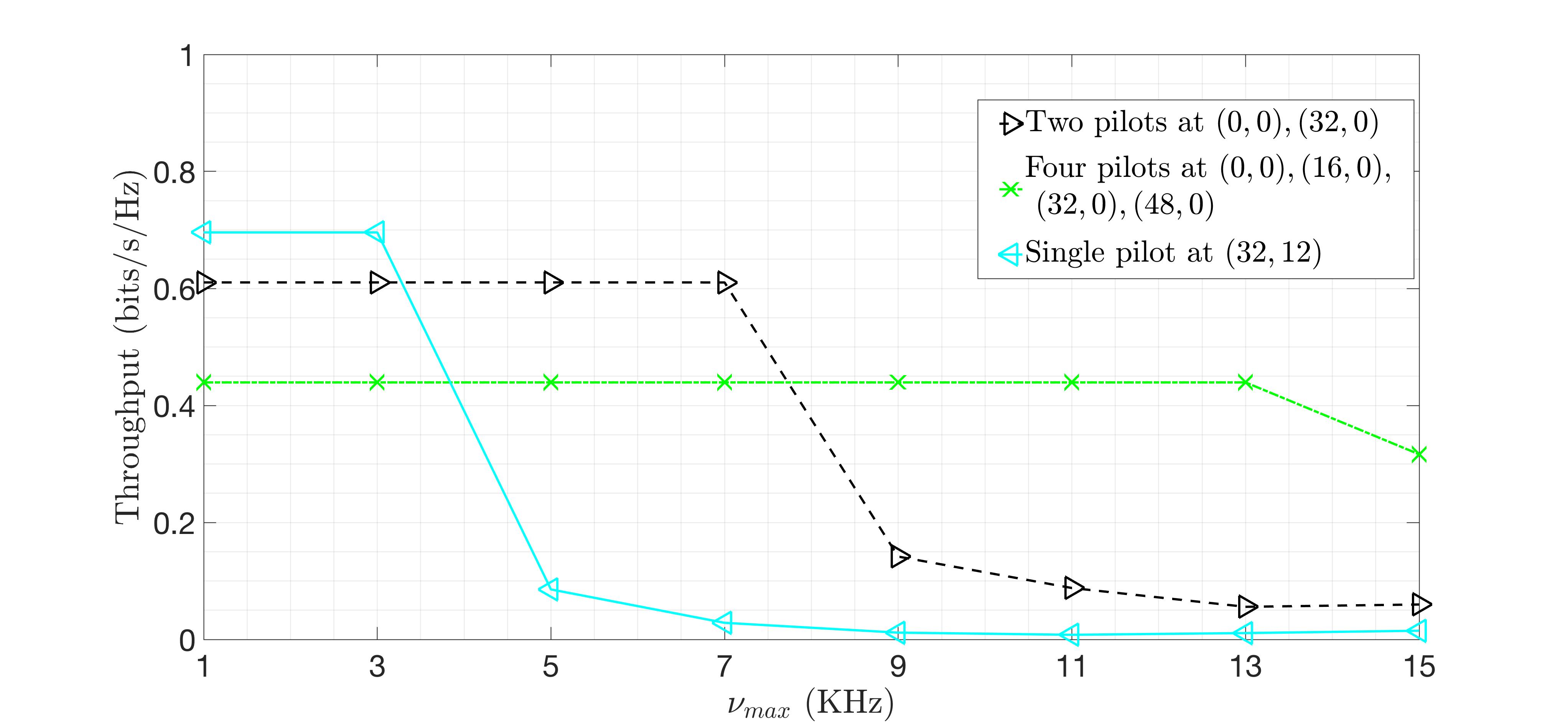}
        \vspace{-2mm}
        \caption{Effective throughput vs. $\nu_{max}$ for the same setting as in Fig.~\ref{fig_bervsnumaxtwopilots}.}
        \label{fig_throughput}
        \vspace{-3mm}
    \end{figure}

In Fig.~\ref{fig_throughput} we plot the effective throughput as a function of increasing $\nu_{max}$ for the same simulation setting as in Fig.~\ref{fig_bervsnumaxtwopilots}. 
Effective throughput is the ratio of the number of bits reliably communicated in each subframe to
the available degrees of freedom ($ B' T'$).
The number of bits communicated reliably in a subframe is simply $(1 - H(BER))$ times the number of information bits transmitted in each subframe. 
Here BER denotes the bit error rate and $H(\cdot)$ denotes the binary entropy function. We maximize effective throughput by interleaving the minimum number of pilots required to accurately estimate
the effective channel. We minimize the number of pilots to avoid introducing unnecessary guard and pilot regions that would reduce effective throughput. When $2 \nu_{max} < \nu_p$ we use a single pilot, when $\nu_p < 2 \nu_{max} < 2 \nu_p$ we
use $2$ interleaved pilots, and when $2 \nu_p < 2 \nu_{max} < 4 \nu_p$ we use $4$ interleaved pilots. Note that although a higher number of interleaved pilots results in stable throughput for a wider range of Doppler spreads (i.e., extension of the region of predictable operation), the throughput achieved is smaller due to a higher pilot and guard region overhead.
{Since $k_{max} = \lceil M \tau_{max} /\tau_p \rceil = 2$, with single, double and four interleaved pilots, the fractional pilot overhead is $7/64$, $14/64$ and $28/64$ respectively.}.

        \begin{figure}
     \centering
        \includegraphics[width=9.1cm,height=6.1cm]{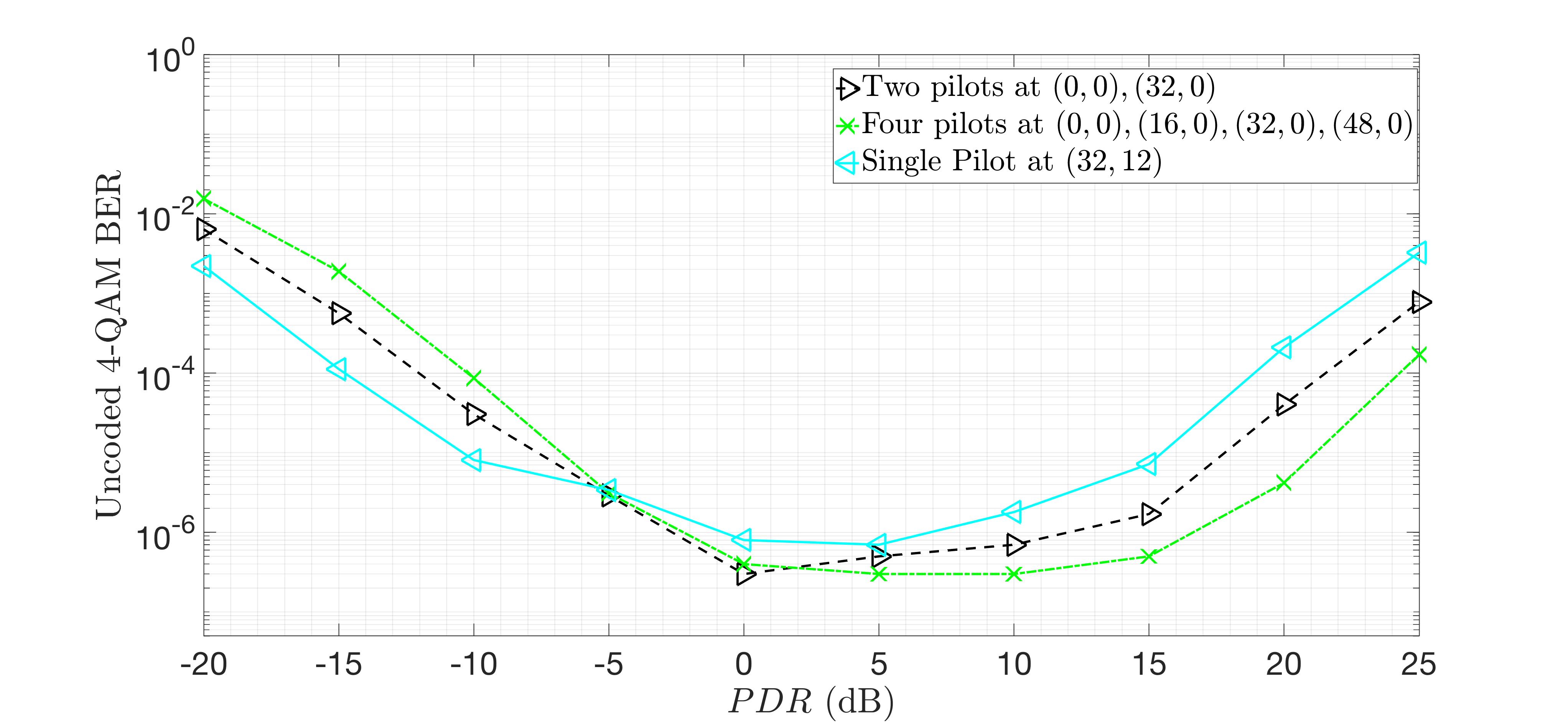}
        \vspace{-2mm}
        \caption{BER vs PDR for six-path Veh-A channel used in Fig.~\ref{fig_bervsnumaxtwopilots} and \ref{fig_throughput}, with a fixed $\nu_{max} = 2.5$ KHz.}
        \label{fig_bervspdr}
        \vspace{-3mm}
    \end{figure}Fig.~\ref{fig_bervspdr} illustrates BER performance as a function of increasing PDR.
    The characteristic ``U" shape is independent of the number of interleaved pilots. At low PDR, estimation of the
    effective channel is inaccurate, hence BER performance is poor.
    As the pilot becomes stronger, effective channel estimation
    becomes more accurate and BER improves. When the pilot power exceeds data power, interference to data from the pilot dominates over noise, and the BER degrades as the PDR increases.

\begin{figure}
     \centering
        \includegraphics[width=9.3cm,height=5.9cm]{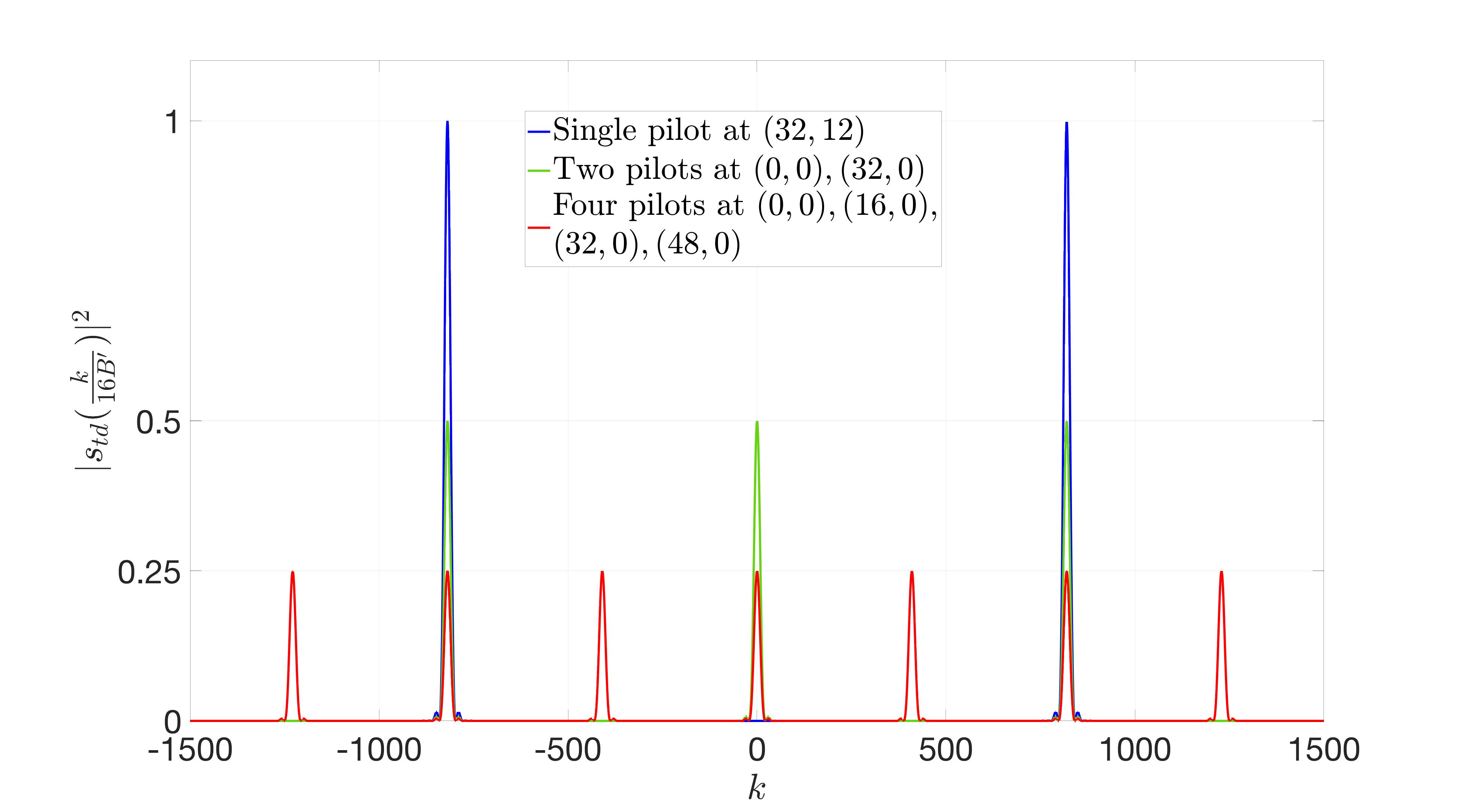}
        \vspace{-2mm}
        \caption{Energy of the TD samples of interleaved pilot samples (sampled at rate $16 B'$) for the Zak-OTFS system
        parameters considered in this Section \ref{simsec}.}
        \label{fig_tdpapr}
        \vspace{-3mm}
    \end{figure}

    The peak to average power ratio (PAPR) of the transmitted TD pilot (no data transmission) depends on the number of interleaved pilots. For the Zak-OTFS system considered here, the PAPR decreases from $19.4$ dB, to $16.4$ dB to $13.4$ dB as the number of interleaved pilots increases from $1$, to $2$, to $4$. This reduction is illustrated in Fig.~\ref{fig_tdpapr}. When the number of interleaved pilots doubles, the separation between
    pulses in the TD pilot pulse train halves.\footnote{\footnotesize{The TD realization of a single impulse pilot at DD location $(k_{p_1}, 0)$ is a TD pulse train with narrow TD pulses at time instances $\left(\frac{k_{p_1} \tau_p}{M} + n \tau_p\right)$, $n=-N/2, \cdots, 0, (N/2 -1)$ for even $N$ (see \cite{zakotfs1} and \cite{zakotfs2}). Therefore, the TD realization of $Q$ interleaved pilots spaced regularly apart at DD locations $(iM/Q, 0)$, $i=0,1,\cdots, (Q-1)$ is a superposition of $Q$ pulse trains, each pulse train being the TD realization to one of the $Q$ pilots. The TD realization of $Q$ interleaved pilots is therefore a pulse train consisting of narrow TD pulses spaced $\tau_p/Q$ seconds apart. For the same total pilot energy $E_p$, the energy of each narrow TD pulse is therefore $E_p/Q$.}}
    There are twice as many pulses, and each pulse is scaled down by $\sqrt{2}$
    to maintain constant average power. Hence the PAPR is halved with every doubling of the number of interleaved pilots.\footnote{\footnotesize{{The PAPR of full signal (comprising both data and
pilot) is smaller than that of a pilot only signal. This is because, pilot is transmitted on only a single pulsone, whereas each data symbol is transmitted on a different pulsone which reduces the peaky nature of the overall signal in TD since data pulsones located on different delay bins have peaks at different time instances. Please see Fig.~$11$ in \cite{zakotfs3}.}}}

{In summary, with increasing number of interleaved pilots, the PAPR decreases, the guard/pilot region overhead increases and the complexity of acquiring the I/O relation also increases.
As long as $\nu_{max} < Q \nu_p/2$, the error rate performance does not floor.}

    \section{Conclusions}
{We have introduced a framework for pilot design in the DD domain which makes it possible to support users with very different delay-Doppler characteristics when it is not possible to choose a single delay and Doppler period to support all users. We have translated the problem of I/O reconstruction to that of designing an interleaved pilot consisting of Zak-OTFS carriers which are selected so that in combination they produce zeros in the auto-ambiguity function of the interleaved pilot. When the interleaved pilots are spaced regularly, the auto-ambiguity function is supported on a sub-lattice of the information grid, and the I/O relation can be reconstructed from the restriction of the cross-ambiguity function (between the received and the transmitted interleaved pilot) to any fundamental region of this sub-lattice. 
Since the nominal complexity of computing the cross-ambiguity
is high, we have introduced a method of estimating the I/O relation that only requires solving a small system of linear equations.}

\appendices
\section{{Proof of Theorem \ref{thm1}}}
\label{prfthm1}
{Substituting the expression of $x_{p,\scriptsize{dd}}[k,l]$ from (\ref{xpddqkl}) into the R.H.S. of (\ref{axpxpkl}) we get (\ref{axpxpkl23}) (see top of next page). With $k'=0,1 \cdots, M-1$, and $l'=0,1,\cdots, N-1$, and $i_1 = 1,2,\cdots, Q$, the Dirac-delta terms $\delta\left[ k' - (i_1 - 1) \frac{M}{Q} - n_1 M\right]$ and $\delta\left[ l' - m_1 N\right]$ are non-zero if and only if $l' = n_1 = m_1 = 0$ and $k' = (i_1 - 1)M/Q$. With this we then get the second step in (\ref{axpxpkl23}). In the R.H.S. of the second step, the term $\delta\left[ - l - m_2 N\right]$ implies that $A_{x_p,x_p}[k,l]$ is non-zero only when $l$ is an integer multiple of $N$ (i.e., $l = -m_2 N$ for some $m_2 \in {\mathbb Z}$). }
\begin{figure*}
{
{
\vspace{-6mm}
    \begin{eqnarray}
    \label{axpxpkl23}
    A_{x_p,x_p}[k,l] & = & \frac{E_p}{Q} \sum\limits_{k' = 0}^{M-1} \sum\limits_{l' = 0}^{N-1} \sum\limits_{i_1 = 1}^Q \sum\limits_{i_2 = 1}^Q \sum\limits_{n_1, n_2, m_1, m_2 \in {\mathbb Z}} \hspace{-1mm} {\Bigg (} e^{-j 2 \pi l \frac{(k' - k)}{MN}} \, \delta\left[ k' - (i_1 - 1) \frac{M}{Q} - n_1 M\right] \, \delta\left[ l' - m_1 N\right] \nonumber \\
    & & \hspace{55mm} \, \delta\left[ k' - k - (i_2 - 1) \frac{M}{Q} - n_2 M\right] \, \delta\left[ l' - l - m_2 N\right] \, {\Bigg )} \nonumber \\
    & = & \frac{E_p}{Q} \sum\limits_{i_1 = 1}^Q \sum\limits_{i_2 = 1}^Q \sum\limits_{n_2, m_2 \in {\mathbb Z}} \hspace{-1mm} e^{-j 2 \pi  l \frac{( (i_1 - 1)\frac{M}{Q} - k) }{MN}} \, \delta\left[ - k + (i_1 - i_2) \frac{M}{Q} - n_2 M\right] \, \delta\left[ - l - m_2 N\right]
    \end{eqnarray}}}
    \vspace{-4mm}
    \begin{eqnarray*}
        \hline
    \end{eqnarray*}
\end{figure*}
{Therefore, for $l = -m_2 N$ the expression for $A_{x_p, x_p}[k,l= -m_2N]$ is given by (\ref{axpxpkl2}) (see top of next page). From the Dirac-delta term in the R.H.S. of (\ref{axpxpkl2}) it follows that $A_{x_p,x_p}[k,l=-m_2 N]$
is non-zero only for $k = (i_1 - i_2)M/Q - n_2M$, i.e., $k$ is an integer multiple of $M/Q$. In general, $A_{x_p,x_p}[k,l=-m_2 N]$ is non-zero only for $k = pM/Q -n_2 M$, where $p =0,1,\cdots, (Q-1)$ and $n_2 \in {\mathbb Z}$. From this, it follows that for any $k \equiv pM/Q$ (modulo $M$) (i.e., $k$ is congruent to $pM/Q$ modulo $M$), we get (\ref{axpxpkl3}).
From (\ref{axpxpkl3}), it is clear that $A_{x_p,x_p}[k,l]$ is non-zero and equal $E_p$ if and only if $k$ is an integer multiple of $M/Q$ and $l$ is an integer multiple of $QN$.}
\begin{figure*}
{
\vspace{-6mm}
{
    \begin{eqnarray}
    \label{axpxpkl2}
    A_{x_p,x_p}[k,l=-m_2 N]
    & = & \frac{E_p}{Q} \sum\limits_{i_1 = 1}^Q \sum\limits_{i_2 = 1}^Q \sum\limits_{n_2 \in {\mathbb Z}} \hspace{-1mm} e^{j 2 \pi  m_2 \frac{( (i_1 - 1)\frac{M}{Q} - k)}{M}} \, \delta\left[ - k + (i_1 - i_2) \frac{M}{Q} - n_2 M\right] 
    \end{eqnarray}}}
    \vspace{-4mm}
    \begin{eqnarray*}
        \hline
    \end{eqnarray*}
\end{figure*}
\begin{figure*}
{
\vspace{-6mm}
{
    \begin{eqnarray}
    \label{axpxpkl3}
    A_{x_p,x_p}[k \equiv pM/Q \,\, (\mbox{\small{mod}} \, M) \, , \, l=-m_2 N]
    & = & \frac{E_p}{Q} \sum\limits_{i_1 = 1}^Q  \hspace{-4mm}\sum\limits_{\substack{i_2 = 1 \\ (i_1 - i_2) \equiv p \, (\mbox{\small{mod}} \, Q)}}^Q  \hspace{-8mm} e^{j 2 \pi  \frac{(i_2 - 1) m_2}{Q}}  \nonumber \\
    & = & \frac{E_p}{Q} \sum\limits_{i_1 = 1}^Q  e^{j 2 \pi  \frac{(i_1 - 1 - p) m_2}{Q}} \, = \, \frac{E_p}{Q} \, e^{-j 2 \pi \frac{p m_2}{Q}} \, \sum\limits_{i_1 = 0}^{Q-1} e^{j 2 \pi  \frac{i_1  m_2}{Q}}  \nonumber \\
    & = & \begin{cases}
          E_p &, \frac{m_2}{Q} \, \mbox{\small{is integer}} \\
          0 &,  \mbox{\small{otherwise}} \\
        \end{cases}.
    \end{eqnarray}}}
    \vspace{-4mm}
    \begin{eqnarray*}
        \hline
    \end{eqnarray*}
\end{figure*}


\end{document}